\documentclass[12pt,preprint]{aastex}
\newcommand{\bmax}{$T(B_{\rm max}$)}
\usepackage{lscape}

\newcommand{\dm}{\Delta m_{15}(B)}
\usepackage{graphicx}
\usepackage{epstopdf}
\usepackage{color}
\usepackage{amssymb}
\usepackage{mathrsfs}  
\usepackage{ulem}
\newcommand{\nosne}{75}

\shorttitle{SNe~Ia NIR $K$~corrections}
\shortauthors{Boldt et al.}

\begin{document}

\title{Near-infrared $K$ corrections of  Type Ia Supernovae and their errors}

\author{Luis N. Boldt\altaffilmark{1},
Maximilian D. Stritzinger\altaffilmark{2},
Chris Burns\altaffilmark{3},
Eric Hsiao\altaffilmark{4},
M.~M. Phillips\altaffilmark{4},
Ariel Goobar\altaffilmark{5},
G. H. Marion\altaffilmark{6,7},
and
Vallery Stanishev\altaffilmark{8}
}

\altaffiltext{1}{Argelander Institut f\"ur Astronomie, Universit\"at Bonn, Auf dem H\"ugel 71, D-53111 Bonn, Germany.}

\altaffiltext{2}{ Department of Physics and Astronomy, Aarhus University, Ny Munkegade 120, DK-8000 Aarhus C, Denmark.}

\altaffiltext{3}{Observatories of the Carnegie Institution for Science, 813 Santa Barbara St., Pasadena, CA 91101, USA.}

\altaffiltext{4}{Las Campanas Observatory, Carnegie Observatories, Casilla 601, La Serena, Chile.}

\altaffiltext{5}{The Oskar Klein Centre, Department of Physics, AlbaNova, Stockholm University, SE-106 91 Stockholm, Sweden.}

\altaffiltext{6}{University of Texas at Austin, 1 University Station C1400, Austin, TX 78712-0259, USA.}

\altaffiltext{7}{Harvard-Smithsonian Center for Astrophysics, 60 Garden St., Cambridge, MA 02138, USA.}

\altaffiltext{8}{CENTRA - Centro Multidisciplinar de Astrofõsica, Instituto
Superior Tecnico, Av. Rovisco Pais 1, 1049-001 Lisbon, Portugal.}

\begin{abstract}
In this paper we use near-infrared (NIR) spectral observations of Type~Ia supernovae
(SNe~Ia) to study the uncertainties inherent to  NIR $K$ corrections. 
To do so, 75 previously published NIR spectra of 33 SNe~Ia are employed to determine $K$-correction uncertainties in the 
$YJHK_s$ passbands as a function of temporal phase and redshift. 
The resultant $K$ corrections are then fed into an interpolation algorithm 
that provides mean $K$~corrections as a function
of temporal phase and robust estimates of the 
associated errors.  These uncertainties are both statistical and intrinsic ---
i.e., due to the diversity of spectral features from object to object --- and must be
included in the overall error budget of cosmological parameters constrained 
through the use of NIR observations of SNe~Ia.  Intrinsic variations are likely the
dominant source of error for all
four passbands at maximum light.  Given the present data, the total
$Y$-band $K$-correction uncertainties at maximum are smallest, 
amounting to $\pm 0.04$~mag at a redshift
of $z = 0.08$.  The $J$-band $K$-term errors are also reasonably small
($\pm 0.06$~mag), but intrinsic variations of spectral features and noise 
introduced by telluric corrections  in the $H$-band
currently limit the total $K$-correction errors at maximum to $\pm 0.10$~mag 
at $z = 0.08$.  Finally, uncertainties in the $K_s$-band $K$~terms at
maximum amount to $\pm 0.07$~mag at this same redshift.  
These results are largely constrained by the small number of published NIR spectra of
SNe~Ia, which do not yet allow spectral templates to be constructed as a function
of the light curve decline rate. 

\end{abstract}


\section{INTRODUCTION}

Type Ia supernovae (hereafter SNe~Ia) are standardizable distance indicators at optical
wavelengths that provide critical constraints on cosmological parameters \citep[see][and references therein]{goobar11}. 
A number of groups have worked diligently to gather optical photometry of homogenous samples of 
low-, intermediate- and high-$z$  SNe~Ia that, when combined, amount to  well over 1000 objects.
As the sample size has increased, 
systematic effects have come to dominate the final uncertainty in 
the measured value of the equation-of-state parameter of the Universe, $w$
\citep[e.g., see][]{wood07,astier06,freedman09,kessler09,folatelli10,conley11,suzuki12}.

A major systematic  that  plagues SN~Ia cosmology 
 is our inability to accurately  estimate host galaxy dust extinction due
 to uncertainties in the  reddening law and, in particular, 
 variations in the value of the total-to-selective extinction, $R_V$, from
 object to object \citep[see][and references therein]{phillips12}. 
 This is further exacerbated by any systematic error in the relative zero points between 
 the nearby and distance SNe~Ia samples, thereby making the latter artificially redder
 (or bluer) than the former.  
 A sensible way around these problems is to observe in rest-frame near-infrared (NIR) bands instead of rest-frame optical bands, because the effects of
dust extinction are minimized and essentially independent 
of the adopted reddening law \citep{krisciunas00}.
 
Additional motivation is provided by empirical evidence indicating that the
luminosities of SNe~Ia show little or no dependence on 
decline rate\footnote[8]{The decline rate of a SN~Ia is 
traditionally defined as the change in its $B$-band magnitude from the time of maximum brightness
to 15 days later, and is denoted as  $\Delta$$m_{15}(B)$.   The decline rate is known
to correlate with the peak absolute luminosity in such a way that more luminous objects exhibit 
smaller  $\Delta$$m_{15}(B)$  values \citep{phillips93}.}  
in the NIR \citep{meikle00,kr204,wood08,kriciunas09,mandel09,folatelli10,kattner12}.
This translates to a reduced intrinsic dispersion in the NIR Hubble diagram, and at the same time evolutionary effects 
as a function of redshift on the progenitor populations are potentially minimized. 
These  factors have provided significant impetus for future SN~Ia cosmology studies to 
construct {\it homogenous samples} of 
low-  and high-$z$  SN~Ia observed in the rest-frame NIR 
\citep[e.g., see][]{green12}.

Reaping the benefits afforded by observing SNe~Ia in the NIR 
requires the development of tools to  
obtain rest-frame luminosities via $K$~corrections \citep{oke68}. 
The $K$~correction is defined as the difference in brightness between an object observed 
 in its rest-frame with a given passband compared to its measured brightness with the same passband when observed at redshift $z$.
Specifically, for a given passband, $i$, the $K$~correction is defined as:
\begin{equation}
m_{i}=M_{i} + \mu + K_{i}.
\label{K_corr1}
\end{equation}
Here $m_{i}$ is  the SN~Ia apparent magnitude observed on Earth, and $M_{i}$ is its absolute magnitude.
The magnitude difference encapsulated in the $K$ term is explained by the shifting and stretching
of an object's spectral energy distribution  (SED), which is inherent to cosmological expansion.

The first calculations of SN~Ia $K$~corrections at optical wavelengths were 
published by \citet{leibundgut90} and \citet{ham93}.  In the latter paper,
optical observations of three nearby SNe~Ia were used to 
construct a sequence of $B$- and $V$-band $K$~corrections extending to a redshift 
of $z = 0.5$. The authors found that the temporal variation of the computed values largely mimicked the 
$(B-V)$ color evolution, implying that the $K$~correction is, to first order, driven by 
the color of the SN.  Shortly thereafter,  \citet{kim96} presented a method to 
compute cross-band $K$~corrections which is  particularly well-suited for SNe~Ia 
located at $z > 0.2$.   \citet{nugent02} expanded upon these efforts by developing  
a set of SN~Ia  optical spectral templates that enabled $K$~corrections to be 
computed as a function of  temporal phase for any given optical bandpass. 
Later, \citet{hsi07} constructed improved optical spectral templates based on a greatly 
expanded sample of nearby SN~Ia.  They found that besides color, spectral diversity 
also has a significant impact upon the magnitude of the $K$~term.
The \citet{hsi07} template is now routinely used to $K$~correct optical photometry of SNe~Ia, 
and the statistical uncertainty associated with these corrections 
is reasonably well understood. 

In comparison, our knowledge of NIR $K$~corrections is still relatively crude.
\citet{krisciunas04} presented the first calculations of the temporal evolution of 
the $K$~term in the $JHK$ passbands based on 11 NIR spectra of SN~1999ee  \citep{hamuy02}.
 Their results indicated that NIR $K$~corrections are non-negligible even at relatively small
 redshifts.
  These initial findings emphasized  the importance of accurately characterizing 
 NIR $K$~corrections {\it and} their uncertainties as a function of redshift and temporal phase.
 Five years later, the publication of 41 spectra by \citet{marion09} revolutionized
 the study of the NIR spectral characteristics of SNe~Ia, and \citet{hsiao09} used these data
 along with the other published spectra available at that time to extend his 
 spectral template to include NIR wavelengths.  In this paper, we refer to this template
 as the ``Hsiao revised template''. 
 
 In the present paper, we expand on this work by using the existing library of 
 published NIR SNe~Ia spectra to determine the uncertainties inherent to using the Hsiao
 revised spectral template to calculate NIR $K$~corrections, particularly those due to intrinsic 
 variations in spectral features. 
 To do this, we first color match\footnote[9]{Often people tend to use 
 the nomenclature warp or mangle, however, we feel that it is more appropriate and accurate to adopted the term color match.}
each observed spectrum to the template, and
 then calculate  $YJHK_s$-band $K$~corrections.
  An interpolation algorithm based on Gaussian Processes combined with 
 Markov-Chain Monte-Carlo methodology is then used to produce mean
 $K$~corrections as a function of temporal phase and redshift,
 along with estimates of both statistical and intrinsic sources of uncertainties.
 These errors, in turn, will provide important input to future studies utilizing the
 NIR light curves of SNe~Ia to estimate cosmological parameters.
 
  The organization of this article is as follows. In
Section~\ref{kcorrection}  the concept of the $K$~correction is briefly reviewed;  
Section~\ref{se:data} introduces the data used in our calculations, 
along with the adopted passbands and methods used to interpolate our computed $K$~corrections; 
Section~\ref{se:results} contains the results; and finally, Section~\ref{conclusions} presents our conclusions.

\section{The $K$~CORRECTION}
\label{kcorrection}
 
In this study we limit ourselves to the discussion of single-band $K$ corrections
rather than cross-band $K$ corrections, which were the subject of papers by 
\citet{nugent02} and \citet{hsi07}.
Given a SED, $f(\lambda)$, and a transmission curve of a particular 
passband, $S_{i}(\lambda)$, the $K$~term 
is computed following Eq.~(\ref{K_corr2}) of \citet{oke68}:
\begin{equation}
K_{i}(z) =2.5 \cdot {\rm log}(1+z)+2.5 \cdot {\rm log}\left[\frac{\int f(\lambda)S_{i}(\lambda) \lambda d\lambda}{\int f(\frac{\lambda}{1+z})S_{i}(\lambda) \lambda d\lambda}\right].
\label{K_corr2}
\end{equation}

\noindent The first ``bandwidth" term of Eq.~(\ref{K_corr2}) is independent of
wavelength and accounts for the 
narrowing of the observed filter as the SED is stretched as a function of redshift
\citep{sandage95}. The second term is (for a given passband) the ratio of the response of 
the SED at the rest wavelength compared to that measured at a redshift $z$,
and accounts for the effects of doppler shifting. 
 This term includes 
 a factor of $\lambda$ in both the numerator and denominator because 
 most modern photometric systems count photons rather than
 energy \citep[for details see][]{nugent02}.

\begin{figure}
\figurenum{1}
\epsscale{1.0}
\plotone{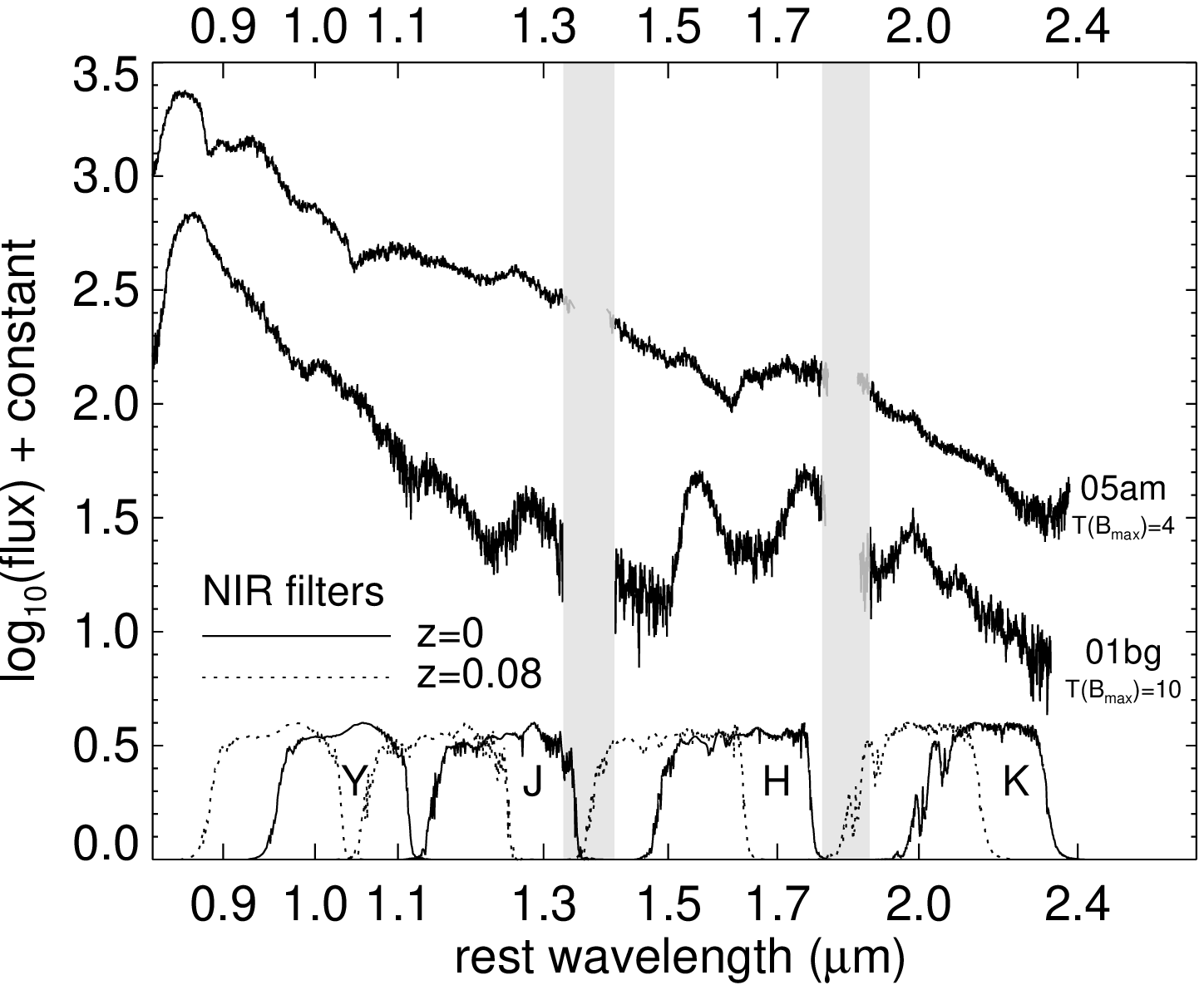}
\caption[]{NIR spectra of the normal Type~Ia  SNe~2005am and 2001bg obtained 
$+$4 and $+$10 days relative to \bmax, respectively. Also plotted as black lines
are the rest-frame CSP natural system NIR transmission functions. In addition,
these passbands are also plotted  as dotted lines at 
the locations which cover the SED they sample when observing a SN~Ia at $z = 0.08$. The vertical bands correspond to regions containing prevalent 
telluric absorption.   
\label{filters}}
\end{figure}

In what follows we adopt for  $S_{i}(\lambda)$ the
$YJHK_{s}$ passbands of  the {\it Carnegie Supernova Project} (CSP; Hamuy et al. 2006).
 Currently the CSP  relies on modeled  NIR instrumental passbands which 
  have been  constructed by multiplying together the   
  factory-measured  filter transmissivities with the 
  transmission functions of two  generic mirror reflections, various optical elements,
a HAWAII--1 1024 $\times$ 1024 pixel HgCdTe detector response curve
and a telluric absorption spectrum \citep{hamuy06}.
 The modeled passbands  are electronically 
 available on the  CSP webpage\footnote[10]{http://obs.carnegiescience.edu/CSP},
 and  are plotted in Figure~\ref{filters}, along with  NIR spectra 
 of the normal Type Ia SNe~2005am and 2001bg obtained $+$4 and 
 $+$10 days past  $B$-band maximum (hereafter \bmax), respectively. This comparison highlights the prevalent emission
 features that coincide with the position of the rest-frame $H$ band,
 which emerge and increase to maximum strength within the first two weeks past maximum.
  To illustrate the effect of cosmological redshift each passband is plotted at rest and at  the positions corresponding
 to the wavelength regions of the SED they sample at $z=$ 0.08.
 
\section{METHODS} 
\label{se:data}

\subsection{NIR Spectral Data}

Our analysis is based on \nosne\ published NIR spectra of 33 SNe~Ia that cover the $YJHK_s$ bands.
 The temporal range spanned is from
$-$14.6  to $+$53.8 days  relative to \bmax.
The majority of the data are drawn from the  \citet{marion09} catalog, which were
 obtained between 2000 and 2005 with the NASA Infrared Telescope Facility
 (IRTF) equipped with SpeX, a medium-resolution, NIR spectrograph and imager.
Additional  spectra include the published sequences of 
  SN~1999ee \citep{hamuy02}, SN~2003du \citep{stanishev07},
  SN~2005cf \citep{gall12} and SN~2011fe  \citep{hsiao13}.
 Table~\ref{list} lists each spectrum sorted by 
 phase along with its date of observation, the telescope used to make the  observations, 
 the redshift of each host galaxy as given by NED, and an estimate of $\Delta$$m_{15}(B)$ for each object.
 When possible, the reported phase of a spectrum 
was estimated with respect to \bmax\, measured from its $B$-band light curve.
 However, there are a  fews cases where  no light curve information is available, we therefore  estimated \bmax\ by cross-correlating an 
 optical spectrum  
 to a library  of SN~Ia spectra using the Supernova Identification (SNID) code 
 (Blondin \& Tonry 2007). 
In  these instances,  the estimated temporal phase with respect to 
 \bmax\  has a realistic   uncertainty of  $\pm$3 days \citep{blondin07}.

  \subsection{Assigning Proper Errors}
  
Before $K$~corrections can be calculated, the errors associated with
the spectra must be quantified.  These arise from two principal 
sources: signal-to-noise of the spectrum, and improperly removed 
telluric features, which are especially important to account for at NIR wavelengths.  In the first case, we used the associated error
spectrum when this was available.  Otherwise, an estimate of
the signal-to-noise as a function of wavelength was made for each spectrum, typically 
amounting to 10\% of the signal.  Due to the fact that this is
uncorrelated noise, its effect on the uncertainty of the $K$~corrections
is relatively small.

Errors in the telluric corrections can arise from several factors, but
typically are due to rapid changes in the water vapor conditions.
Also, for some features, the absorption reaches 100\% of the
continuum.   These errors become important at higher redshifts
when the features are shifted into the filter bandpasses.
Unlike the errors due to signal-to-noise, these telluric 
correction errors can be highly correlated if improperly subtracted.  
Fortunately, 
most spectra showed no sign of any systematic over- or under-subtraction, in which case we treated the error as white noise.
 Only in the cases where removal of
the features left a clear imprint on the corrected spectrum did we
simulate an additional error using a telluric template spectrum.

These two errors were then propagated using Monte Carlo simulations.
In brief, 100 artificial spectra were created by introducing noise at a
level consistent with the dispersion of the above-described errors, resulting in a sample of 
100 $K$~corrections.  The standard deviation was then computed 
and adopted as the final error in
the $K$~correction for the particular redshift being calculated.
 
 \subsection{Extending the Wavelength Coverage of the Input Spectra}

The goal of this paper is to quantify the errors in the $K$~corrections due to
variations in {\it spectral features} \citep[cf.][]{hsi07}.  Our approach is to color-match the sample 
spectra to observed colors derived from the Hsiao revised template at a range
of redshifts.  The procedures used for the color-matching are those described by
\citet{hsi07} and \citet{burns11}.

Ideally, when calculating $K$~corrections for a particular supernova, multiple observed 
colors will be available spanning an extended wavelength range so as to properly anchor the
color-matching function.  However, in the present paper, we are dealing with spectral
observations with limited wavelength coverage.  Hence, it is necessary to extend
the sample spectra to bluer and redder wavelengths.  First, for each observed spectrum,
the Hsiao revised template is color-matched to synthetic colors
calculated from the observed spectrum.
  The color-matched template spectrum was then scaled to match the integrated flux under 
  the bluest (or reddest, in the case of the $K_s$ band) 
  filter covered by the observed spectrum.  The template and observed spectra are
  then combined utilizing an overlapping 300 \AA\  region with appropriate weighting.
 Note that the extended sections are not used in the $K$-correction calculations, 
 but serve only for color matching.
 
\begin{figure}
\figurenum{2}
\epsscale{1.0}
\plotone{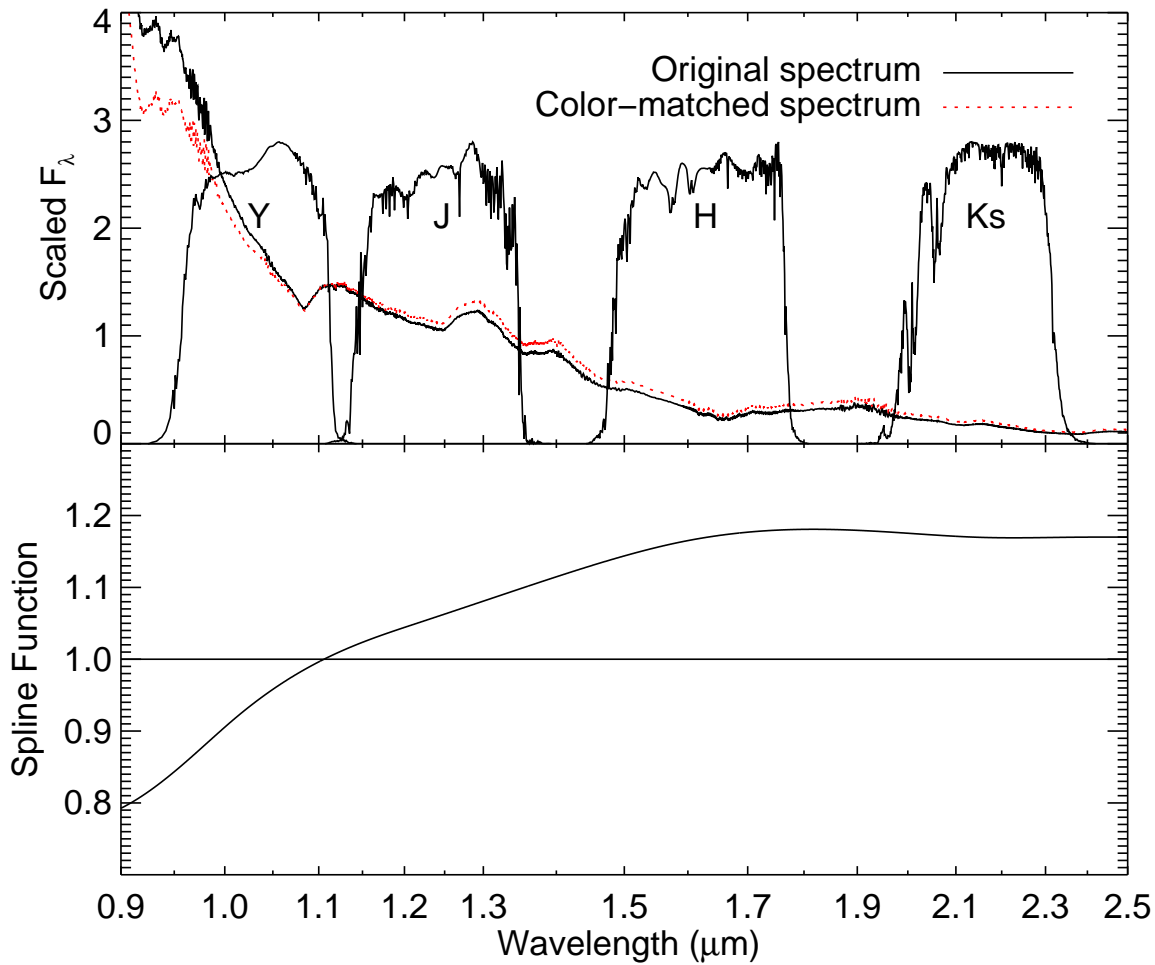}
\caption[]{{\it (top~panel)} Comparison of the maximum light spectrum of SN~2011fe (solid black line) to 
its altered version (dotted red line) that is matched to the 
observed colors derived from the Hsiao revised spectral template at $z = 0.03$.
Overplotted in the top panel are the CSP $YJHK_s$ passbands at rest wavelength.
 {\it (bottom panel)} The spline function used to perform the color matching of the input spectrum to match
the colors calculated from the Hsiao revised spectral template. 
\label{spline}}
\end{figure}

 An example of an original  and color-matched spectrum of a normal SN~Ia that has been extrapolated in the blue is shown in Figure~\ref{spline}.
    
   \subsection{Gaussian Process and MCMC modeling}
 \label{MCMC}
 
 We have adopted a Bayesian formalism to compute an interpolated  function of the 
 $K$~term as a function of light-curve phase, redshift, and filter, along with a robust estimate 
 of its uncertainty, which is more accurate than what is obtained by just simply binning residuals, particularly in the presence of outliers.  
In this manner the computed error snake is a combination of  (i) observational error, (ii) the differing number of points constraining the interpolator at any particular phase and redshift, and  (iii) intrinsic differences between individual SN~Ia. These last two errors are particularly difficult to disentangle, prompting  us to  opt for a combined Gaussian Process and Markov-Chain Monte-Carlo (MCMC) method to obtain an empirically-based model and an associated error estimate. 

 Gaussian Processes \citep{rasmussen06}  are a generalization of probability distributions for functions.  Instead of a mean value, one solves for a mean function, $F(t)$, and 
 in addition,  a covariance function, [$C(t_1,t_2)$]. 
 The covariance function is  a generalization of the standard deviation of a Gaussian distribution and encapsulates the uncertainty in the mean function.
While the mean function is non-parametric, we elected to use the 
Mat\'{e}rn covariance function \citep{banerjee03}, which has four parameters:  scale ($s$), amplitude  ($\theta$), the degree of 
differentiability ($d$), and an intrinsic variance ($\sigma_{var}$).
 The value of $s$ sets the scale over which departures from the mean function are coherent, while $\theta$ determines the size of these departures, $d$ controls the degree of differentiability (or smoothness) of the function, and   $\sigma_{var}$ 
 characterizes any additional uncorrelated dispersion of the interpolated function that exceeds 
 the uncertainty of the $K$~corrections. 
  To implement this approach,  the python package PyMC \citep{patil10} was used to compute  the best values for $s$, $\theta$, $d$, and 
  $\sigma_{var}$, as well as $F(t)$ via the  MCMC method.  

The advantage of MCMC modeling is that it allows one to marginalize over the adopted 
assumptions and missing information, and at the same time provides a robust error snake.
At each MCMC step, $s$, $\theta$ and $d$ are proposed and a Gaussian Process is computed which gives an interpolation between the observed
$K$~corrections. The residuals between the observed points and the interpolated function are then determined. 
 If the residuals are consistent with the errors of the individual $K$~corrections, then  the MCMC process has found a higher likelihood state.  If the residuals are larger than the errors, a lower likelihood state is obtained. 
The Metropolis-Hastings algorithm is used to propose each new MCMC step,
randomly walking through parameter space, though converging to higher probability 
states. 
There are two states where the likelihood is locally maximized:  (i) $s$ and $d$ are small, $\theta$ is large, and the interpolator fits the noise; and (ii) $s$ and $d$ are larger, $\theta$ smaller, and the interpolator smoothly fits the data, though with larger scatter than the errors in the data should allow. Introducing the extra noise term $\sigma_{var}$ gives the second case a higher probability and also gives us an estimate of the intrinsic scatter in the $K$~corrections.

Once the Markov chain has converged, all the states are tallied, and 
 for each  state, $i$, the interpolating function, $F_{i}(t)$, is computed. The 
 median of all $F_{i}(t)$ represents the final $F(t)$, while    
the standard deviation of the interpolations gives the ``error in the mean" function, which we refer to in the following as $\sigma_{stat}$.  
Due to  the varying density of data points used to constrain the interpolation, the value of  $\sigma_{stat}$ varies with both redshift and temporal phase.  The  computed value of  $\sigma_{var}$  is held constant for each
 redshift bin since,
 given the size of the current library of NIR spectra, there simply is not enough data to 
obtain an estimate as a function of temporal phase.
We are unable to directly attribute $\sigma_{var}$  to any one variable, but this 
parameter most likely reflects intrinsic SN-to-SN differences such as 
spectral line diversity. 
It is  our view that a combination of both noise terms   
 ($\sigma_{stat}$ and $\sigma_{var}$)  represents the most
 appropriate error snakes to adopt if one intends to  include the $K$-term uncertainty in the overall error budget of cosmological parameters derived from NIR observations of SNe~Ia.
   
 \section{RESULTS}\label{se:results}

Armed with  the $YJHK_s$ passbands shown in Figure~\ref{filters}
and our adopted library of NIR SN~Ia spectra (see Table~\ref{list}), 
$K$~corrections were computed using Equation~(\ref{K_corr2}) for
 redshifts  of $z = 0.03, 0.05, 0.08$. 
The  arrays of $K$~corrections for each of these redshift intervals 
were then fed into our MCMC algorithm,
from which  smooth mean functions were computed.
The definitive $K$-correction interpolations and their associated values of $\sigma_{stat}$ are provided in Table~2.
Listed in  Table~3 are the computed values of 
$\sigma_{var}$, which dominate the total errors for all four 
passbands near maximum light.
 
\begin{figure}[t]
\figurenum{3}
  \epsscale{1.1}
 \plottwo{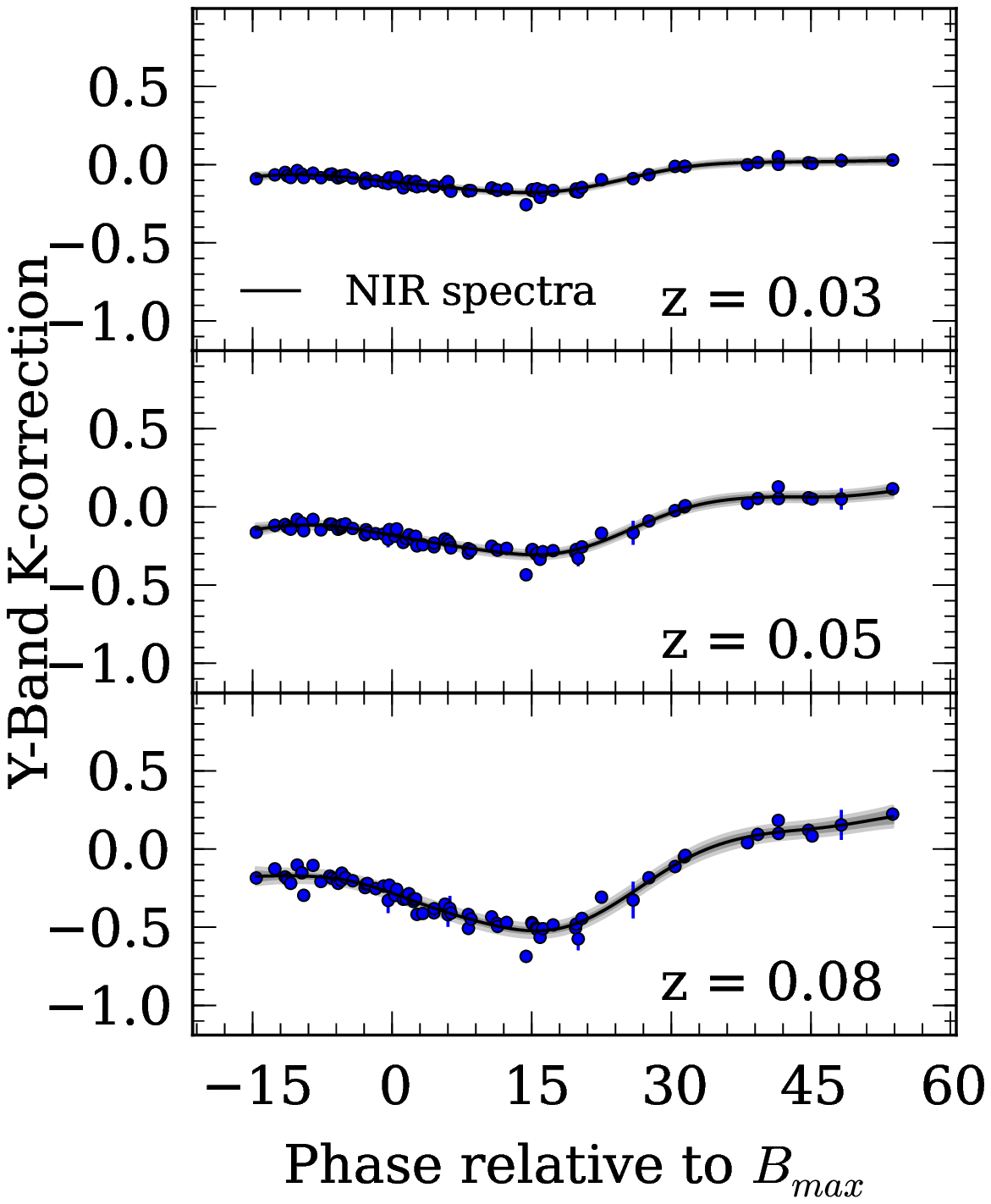}{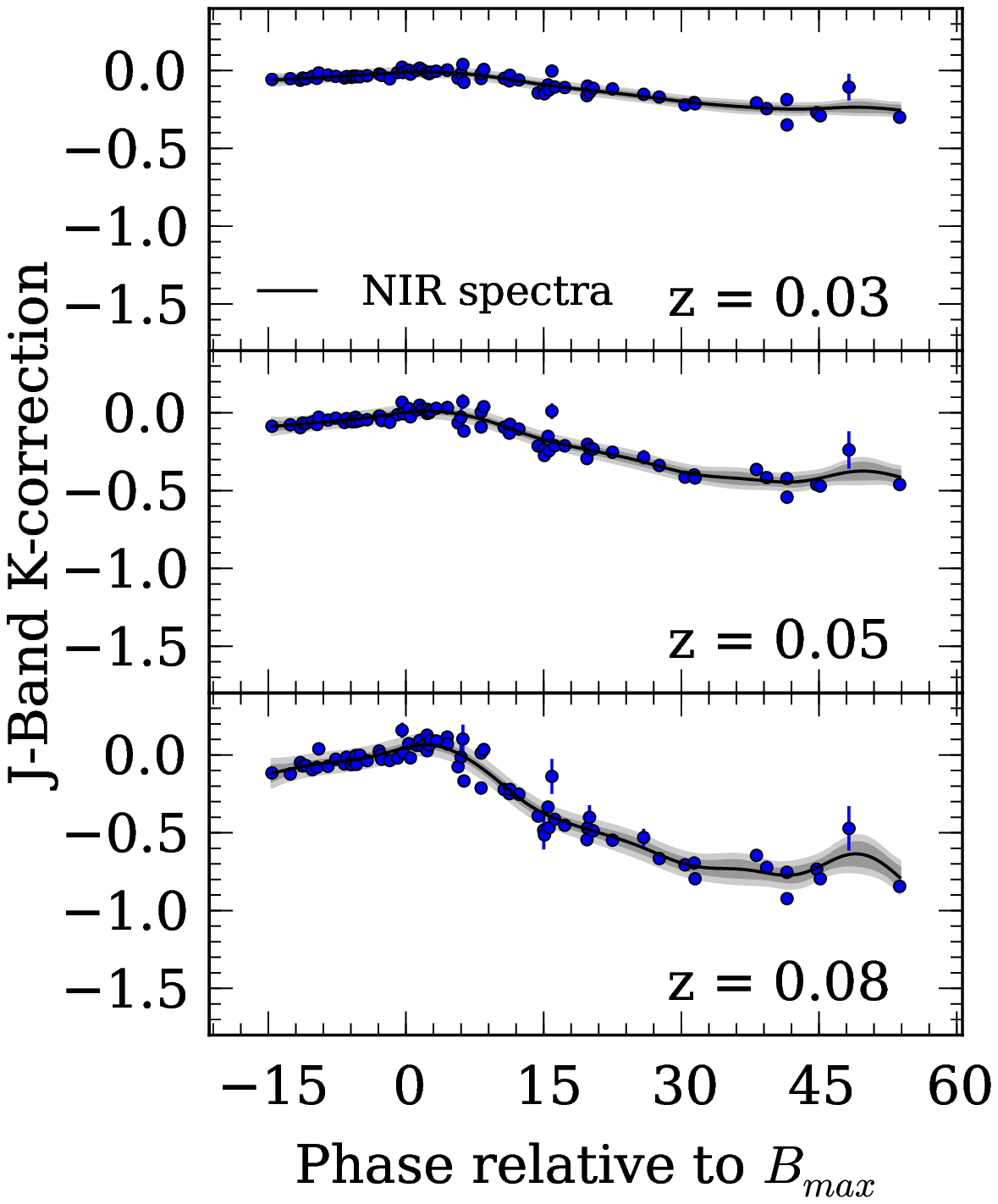}
  \newline
   \newline
 \plottwo{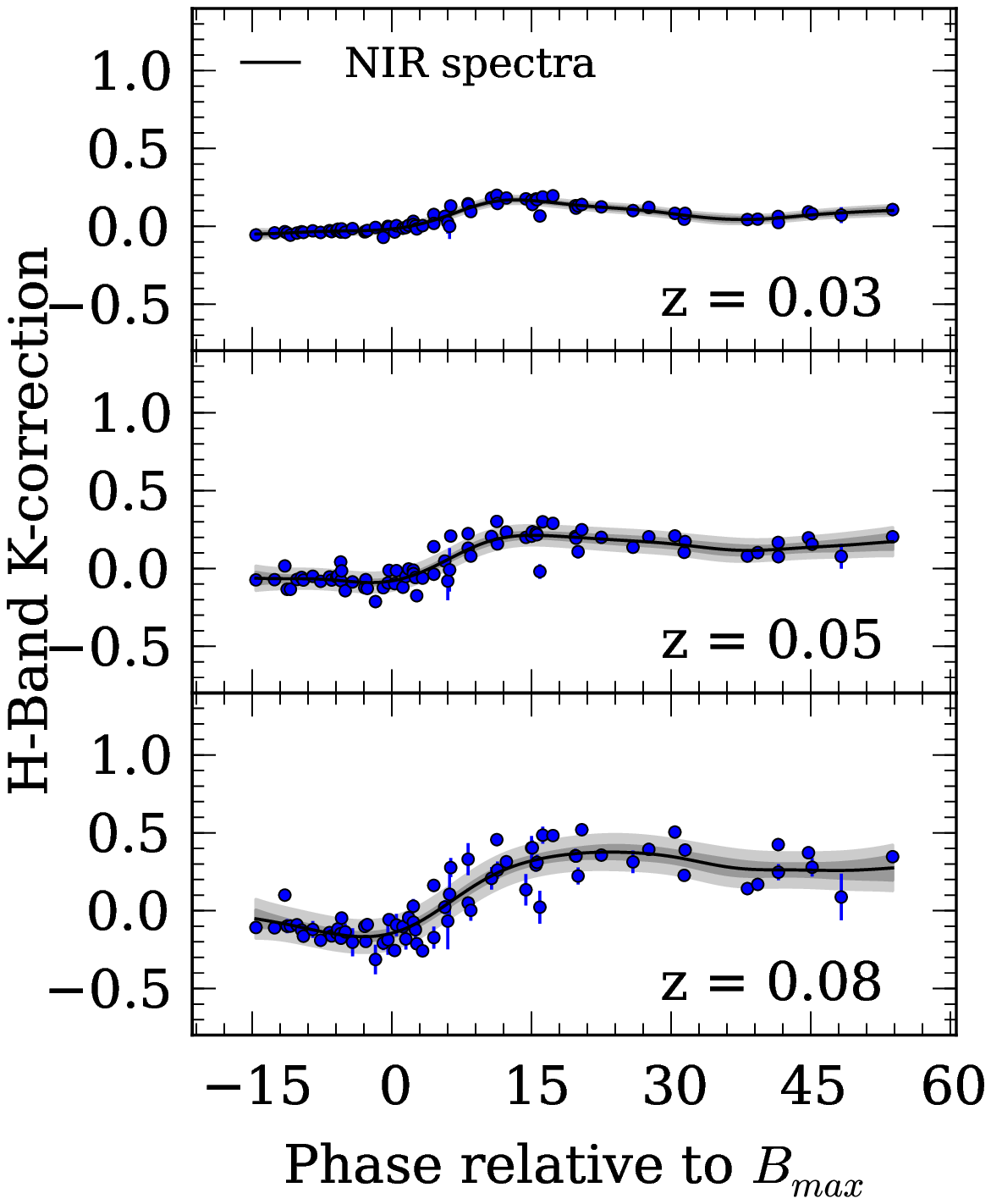}{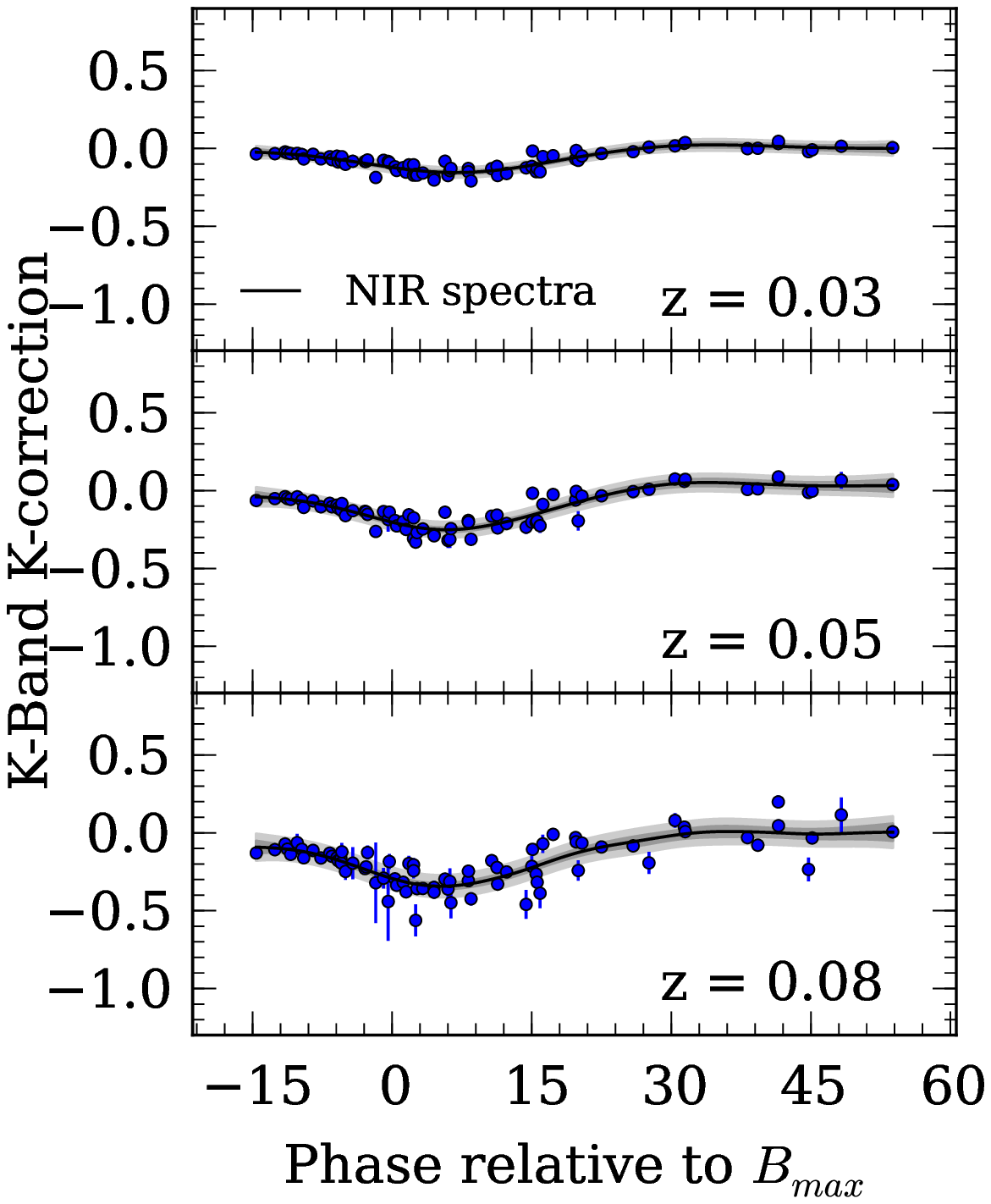}
\caption[]{$K$~corrections in the $YJHK_s$ bandpasses (blue points) computed from \nosne\  
NIR SN~Ia  spectra at  $z = 0.03$, 0.05, 0.08.
Solid black line corresponds to MCMC interpolated functions, while 
dark and light shaded regions correspond to $\sigma_{stat}$, and the summation in quadrature of 
 $\sigma_{stat}$  and $\sigma_{var}$, respectively.\label{kcorrections}
}
\end{figure}

Illustrations of the computed $K$~corrections and their mean interpolated functions 
as a function of temporal phase are presented in Figure~\ref{kcorrections} for the three adopted redshifts.   
In each panel the points correspond to the individual $K$~corrections derived 
from Equation~(\ref{K_corr2}).  The solid black lines  are the  MCMC mean functions, 
and the shaded regions are confidence levels  of these functions. 
 The darker of the two shaded regions corresponds to $\sigma_{stat}$, while the less 
 dark regions are obtained through the summation in quadrature of 
  $\sigma_{stat}$  and $\sigma_{var}$.
  As expected, the overall evolution of the $K$~terms in $Y$, $J$, and $H$ as a function of phase,
mimic to first order the evolution of 
the $(Y-J)$, $(J-H)$, and $(H-K)$ color curves.

We would emphasize that the $K$~terms given in Table~2 are listed for reference
only, and should not be used blindly to correct NIR photometry for any given SN.
Rather, we recommend a two-step process. 
The chosen spectral template 
template should first be color-matched to the SN photometry and then used to 
calculate $K$~corrections for the filter bandpasses employed.  The 
uncertainties given in Tables~2 and 3 should then be interpolated to the redshift of 
the SN and propagated as part of the $K$-correction process.  As mentioned in
\S\ref{MCMC}, we recommend summing in 
 quadrature the values of  
 $\sigma_{stat}$  and $\sigma_{var}$ to estimate the total error in the
 $K$~correction.
  
In the remainder of this section, we discuss results for each of the four CSP filter bandpasses.

   \subsection{$Y$ and $J$ Filters}
  
\begin{figure}[t]
\figurenum{4}
  \epsscale{1.}
 \plotone{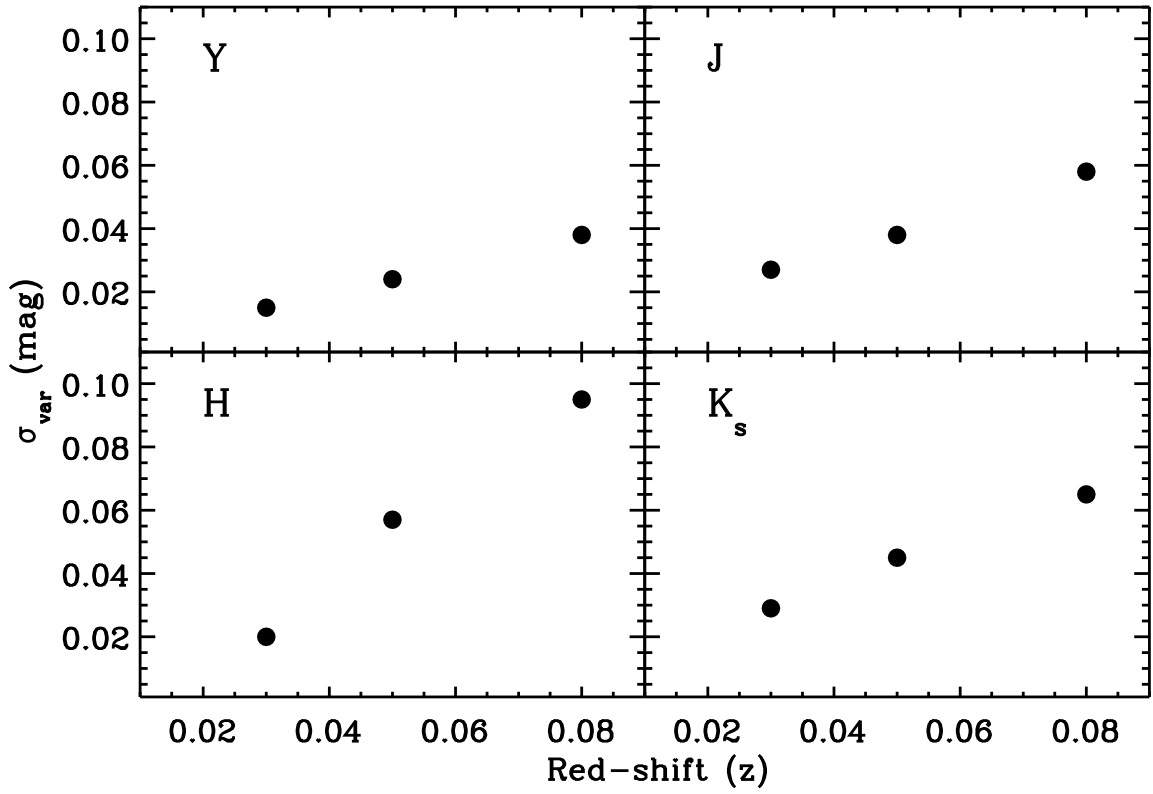}
  \newline
\caption[]{Plot of the uncertainties due most likely to spectral diversity, $\sigma_{var}$, versus redshift. 
   \label{error2}}
\end{figure}

  Figure~\ref{kcorrections} shows that, of the four CSP bandpasses, $Y$ 
  yields the lowest $K$-correction uncertainties.  This is due in large part to 
  the fact that this wavelength region does not develop any strong spectral features
  until $\sim$30~days after \bmax\ \citep[see Figure 9 of][]{marion09}.
  This is reflected in the relatively small increase $\sigma_{var}$ from
  $\pm 0.02$ to $\pm 0.04$~mag between 
  $z = 0.03$ and 0.08 (see Table~3 and Figure~\ref{error2}).  
  Moreover, the $Y$~band is not affected by strong telluric absorption, and is
  generally characterized by higher signal-to-noise than any of the other NIR
  bandpasses.  The small $K$-correction uncertainties combined with
  an essentially negligible dependence of absolute magnitude on decline rate
  \citep{kattner12} and low sensitivity to dust extinction make the $Y$~band
  an interesting option for future cosmological studies.
  Note, however, that to obtain the highest precision in the $Y$ band, it is important that 
  the photometric coverage of the SN extend to at least the $i$ (or $I$) filter
  since this is essential for proper color-matching of the spectral template.
    
  The $K$-correction models indicate that the $J$ passband is also characterized by
  relatively small uncertainties, although not quite at the low levels observed for
  the $Y$~band.  The statistical errors of the $J$-band $K$~terms are slightly 
  higher than those for $Y$ and, as indicated in Table~3 and illustrated in 
  Figure~\ref{error2}, the uncertainties due to intrinsic spectral variations are
  also $\sim$50\% greater.  Nevertheless, the $J$ filter clearly offers considerable 
  promise for cosmological studies.

   \subsection{$H$ Filter}
  
  \citet{kasen06} has argued from theoretical grounds that the spread in the
  absolute magnitudes of SNe~Ia should decrease steadily from the optical to
  the NIR, reaching a minimum dispersion of $\sim \pm0.1$~mag at $H$, and
  \citet{wood08} concluded the same based on observations of 21 SNe~Ia in
  in $JHK$. 
  Nevertheless, Figure~\ref{kcorrections}
  shows that the dispersion in the $H$-band $K$~corrections increases much more
  sharply with redshift than for the $Y$ and $J$ filters, with the error snakes
  indicating that the increases occur in both $\sigma_{stat}$ and 
  $\sigma_{var}$.
  The increase in $\sigma_{stat}$ is predictable since the library spectra
  all correspond to low-redshift ($z \lesssim 0.03$) SNe.  As illustrated in
  Figure~\ref{filters}, the wavelength region of the library spectra affected by the 
  telluric absorption band at 1.34-1.41~$\mu$m is shifted more and more 
  into the blue edge of the $H$ passband at increasing redshift, introducing
  increasing levels of noise into the $K$-correction calculation due to errors 
  in the telluric corrections.  These statistical errors can be beaten down as NIR
  spectra are obtained of SNe~Ia at a larger range of redshifts.  
  
  As shown in Figure~\ref{error2}, the increase in $\sigma_{var}$ with redshift 
  for the $H$~band is greater than that suffered by any of the other CSP NIR filters, 
  reaching a value of nearly $\pm 0.1$~mag at $z = 0.08$.  This behavior is most likely 
  the result of  intrinsic diversity in the strong
  iron-peak emission features that strengthen dramatically in the $H$~band beginning
  only a few days after \bmax\ \citep{wheeler98}. \citet{hsiao13} recently studied
  the strength of the ``break'' at $\sim$1.5~$\mu$m, and found evidence that it
  correlates with $\dm$.  If confirmed, it should be possible to eventually decrease 
  $\sigma_{var}$ by creating NIR spectral templates in the $H$~band that are a 
  function of $\dm$.  This, however, will require a much larger library of NIR
  spectral observations.

  \subsection{$K_s$ Filter}
  
  In principle, the $K_s$ bandpass also offers significant advantages for measuring
  cosmological distances.  Dust extinction is lowest at this wavelength, and the
  increase of $\sigma_{var}$ with redshift is only $\pm 0.01$~mag greater
  than derived for the $J$~band (see Table~3 and 
  Figure~\ref{error2}).  However, the signal-to-noise
  typically achieved in the $K_s$~band is less than in the other NIR
  filters \citep[see Figure 9 of][]{marion09}.  This is due to both the faintness of
  the SNe and the much stronger sky emission encountered
  at these longer wavelengths.  Secondly, similar to the $H$ band, $K$-correction
  calculations for the $K_s$ filter suffer from redshift-dependent noise introduced by the strong telluric
   absorption band at 1.8-1.9~$\mu$m (see Figure~\ref{filters}) although, again, it should
  be possible to minimize the impact of this effect as NIR spectra
  of SNe~Ia are obtained over a larger range of redshifts.  A final 
  disadvantage of the $K_s$ bandpass is the difficulty of properly 
  color-matching it to a spectral template since observations are not generally
  available in a longer-wavelength
  filter to help anchor the correction.

\section{CONCLUSIONS}
\label{conclusions}

Using a library of publicly available SNe~Ia NIR spectra, we have computed 
a set of  $K$~corrections in the CSP $YJHK_s$ filters at redshifts of  $z =0.03$, 0.05, and 0.08.
The individual spectra were first color-matched to the Hsiao revised spectral
template before calculating the $K$~corrections.  
A combined Gaussian Process and MCMC method was then employed to 
derive an empirically-based model of the $K$~terms 
as a function of temporal phase and redshift.  This procedure
returns uncertainties in the $K$-correction model due to both statistical noise
and intrinsic diversity in the features of the input spectra.

\citet{krisciunas04} published a table of  NIR $K$ corrections computed using spectra of SN~1999ee,  and a set of filters functions similar to those adopted in this 
study. A comparison between the results of \citeauthor{krisciunas04}  to those obtained with our expanded set of spectra and more advanced color-matching technique provides 
good agreement in the $J$ and $H$ bands at $z=0.03$, while for the $K_s$ band 
we find differences of up to $\pm$0.08 mag. The discrepancy in the $K_s$ is not surprising owing the the dearth of spectra included in the work of 
\citeauthor{krisciunas04}, as well as the difficulties highlighted in this study 
concerning the difficulties inherent to computing robust $K_s$-band $K$ corrections.

Our emphasis in this paper has not been to provide 
``lookup" tables of NIR $K$~corrections, but rather to 
derive  accurate estimates of the uncertainties inherent in the use
of  spectral templates (like the Hsiao revised template) constructed 
from the NIR sample of SNe~Ia spectra available today.  We
find that the $Y$ and $J$ bands currently afford the greatest precision
in $K$-correction calculations due to the general weakness of the
spectral features and the minimal effect of
telluric corrections.  The $Y$ band is particularly noteworthy,
with the uncertainty due to spectral feature variations increasing from only
$\pm0.02$ to $\pm$0.04~mag from $z = 0.03$--0.05.
The $H$ band is currently more problematic due to noise
introduced by telluric corrections and the appearance just
after maximum light of strong Fe-peak emission features.
Diversity in the strength of this emission most likely explains the strong increase we find in the
intrinsic component of the $K$-correction error from 
$\pm0.02$ to $\pm$0.10~mag over the redshift range $z = 0.03$--0.05.
This result  is a departure from 
pervious theoretical \citep{kasen06} and observational 
\citep{wood08} findings, which have identified the $H$ band as having potentially the least dispersion among the NIR passbands. 
Finally, the uncertainties in the $K$~terms for the $K_s$ filter due
to intrinsic spectral diversity are similar to those found for the $J$~band,
but the currently-available library of spectra covering this
wavelength region suffer from both low signal-to-noise and 
telluric absorption.

\begin{figure}[t]
\figurenum{5}
  \epsscale{1.}
 \plotone{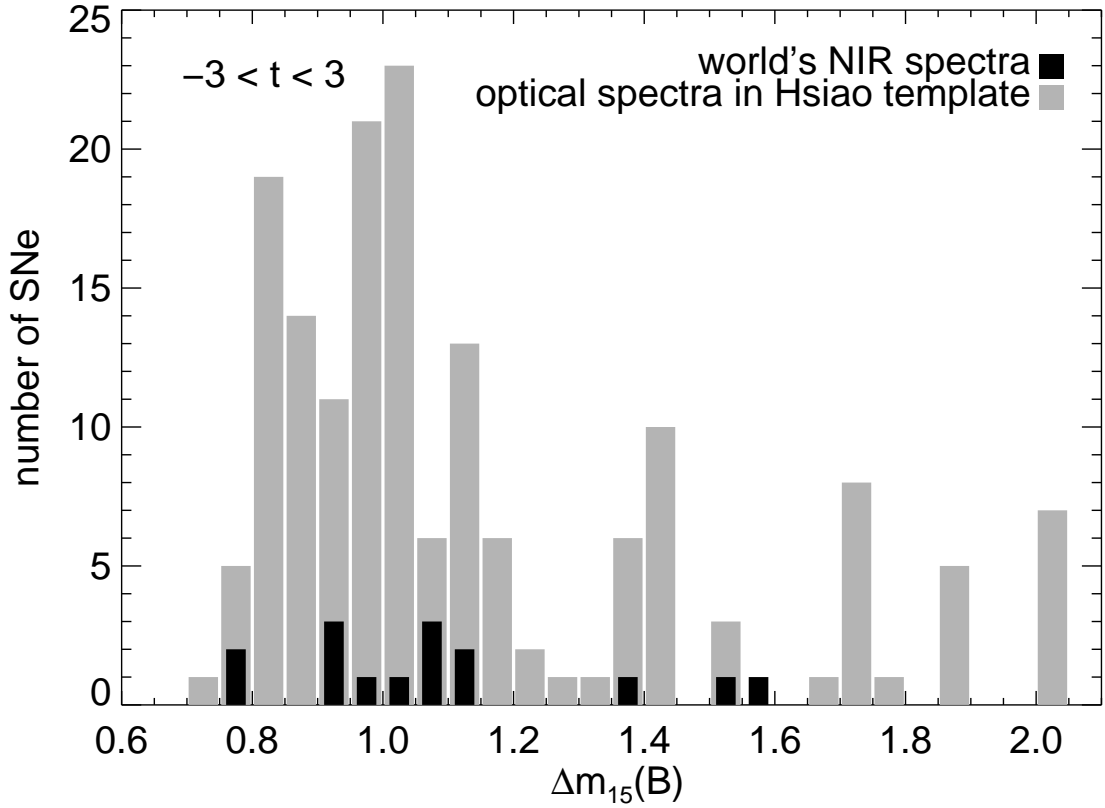}
  \newline
\caption[]{Histogram of the decline rates of the published NIR spectra of SNe~Ia in our sample 
obtained within $\pm3$ days of \bmax.  
A comparable histogram for the optical spectra in the 
\citet{hsi07} template over the same range of epochs is shown for comparison.\label{hist}}
\end{figure}

Further progress in reducing the uncertainties in the 
$K$~corrections for NIR filter bandpasses demands many more
spectral observations over a wider range of redshifts. Figure~\ref{hist}
shows a histogram of the decline rates of the SNe~Ia in our sample
with temporal phases within $\pm 3$~days of  \bmax.  For reference,
the same figure includes a plot of the decline rates
of the spectra employed by \citet{hsi07} to create their optical
spectral template.  The discrepancy between these two histograms
dramatically illustrates the limitations of the currently-available
library of NIR spectra.   
To address this problem, we have embarked on a four-year, 
six-months-a-year SN~Ia followup program that builds upon the 
legacy of the CSP. This project, designated CSP-II,  is designed to obtain 
optical and NIR  light-curves of $\sim$100 SNe~Ia in the smooth Hubble flow
($0.03 < z < 0.08$).  Complementary to these observations, frequent NIR spectroscopy
is being carried out of nearby SNe~Ia, mainly with the Folded-port InfraRed Echellette (FIRE)
spectrograph mounted on the 6.5-m Magellan Baade telescope. 
  Particular emphasis is being placed on 
obtaining spectral sequences for a sub-sample of the SNe that span a range in 
decline rate and phase, which are
essential to determining correlated errors in the $K$~corrections.
We are confident that such observations will not only allow us to more precisely 
characterize NIR $K$-correction uncertainties, but also to decrease 
these errors to the low levels now obtained at optical wavelengths.
 
 In order to ascertain the contribution of NIR $K$-correction errors to the final error budget of cosmological parameters one must determine whether  they are systematic in temporal phase, and therefore propagate to the peak magnitude of a SN~Ia. On the other hand, if the $K$-correction errors are random in phase then the final uncertainty in the peak magnitude will be reduced by template fitting. To determine which of these is the case is, however, beyond the scope of this paper due to the lack of a statistical sample of SNe~Ia with good spectral time series. 
 However, in the near future we will be able to tackle this issue with the high-fidelity spectral series currently being obtained.  

\acknowledgments
M. D. S.  gratefully  acknowledges  generous support provided by the Danish Agency for Science and Technology and Innovation  realized through a Sapere Aude Level 2 grant.
We also acknowledge support from the USA's NSF through grants AST--0306969, AST--0607438 and AST--1008343.  This research has made use of the NASA/IPAC Extragalactic Database (NED) which is operated by the Jet Propulsion Laboratory, California Institute of Technology, under contract with the National Aeronautics and Space Administration.

\clearpage

\clearpage

\clearpage
\newpage

\clearpage
\newpage

\clearpage
\newpage

\clearpage
\begin{deluxetable}{lrcccc}
\tabletypesize{\footnotesize}
\tablenum{1}
\tablecolumns{6} 
\tablewidth{0pc}
\tablecaption{Seventy-five NIR SN~Ia spectra sorted by phase\label{list} }
\tablehead{
\colhead{SN}             &      
\colhead{Epoch\tablenotemark{a}}     &
\colhead{Obs. Date}      &
\colhead{Telescope}      &
\colhead{Redshift}       &
\colhead{$\Delta m_{15}$\tablenotemark{b}} \\
\colhead{}               &
\colhead{$T(B_{max})$}   &
\colhead{(UT)}           &
\colhead{}               &
\colhead{Host ($z$)}     &
\colhead{($B$)} 
}
\startdata
2011fe & $-$14.6(0.5) & Aug. 26.3 & IRTF  &  0.0008 &  1.07$^{8}$     \\
2011fe & $-$12.6(0.5) & Aug. 28.3 & Gem-N &  0.0008 &  1.07$^{8}$     \\
2003du & $-$11.5(0.5) & Apr. 25.0 & UKIRT &  0.0064 &  1.02$^{6}$     \\ 
2002fk & $-$11.3(0.5) & Sep. 19.4 & IRTF  &  0.0071 &  1.08$^{3}$    \\ 
2003du & $-$10.9(0.5) & Apr. 25.9 & TNG   &  0.0064 &  1.02$^{6}$     \\
2005cf & $-$10.2(0.5) & Jun. 01.1 & VLT   &  0.0065 &  1.12$^{9}$    \\
2011fe & $-$09.7(0.5) & Aug. 31.2 & Gem-N &  0.0008 &  1.07$^{8}$     \\
2005cf & $-$09.5(0.5) & Jun. 03.1 & TNG   &  0.0065 &  1.12$^{9}$     \\
1999ee & $-$08.5(0.5) & Oct. 09.0 & VLT   &  0.0114 &  0.94$^{7}$       \\ 
2004bw & $-$07.7(0.5) & May  30.4 & IRTF  &  0.0214 &  1.31$^{3}$    \\ 
2011fe & $-$06.7(0.5) & Sep. 03.2 & Gem-N &  0.0008 &  1.07$^{8}$     \\
2003W  & $-$06.5(0.5) & Feb. 02.4 & IRTF  &  0.0201 &  1.16$^{5}$      \\ 
2002cr & $-$05.9(0.5) & May  08.4 & IRTF  &  0.0096 &  1.26$^{5}$      \\ 
2000dn & $-$05.8(0.5) & Oct. 01.5 & IRTF  &  0.0320 &  1.12$^{3}$    \\ 
2003du & $-$05.5(0.5) & May  01.0 & UKIRT &  0.0064 &  1.02$^{6}$     \\ 
2003W  & $-$05.5(0.5) & Feb. 03.4 & IRTF  &  0.0201 &  1.16$^{5}$      \\ 
2005cf & $-$05.4(0.5) & Jun. 06.1 & VLT   &  0.0065 &  1.12$^{9}$    \\
2004bv & $-$05.0(0.5) & May  30.5 & IRTF  &  0.0105 &  0.89$^{3}$    \\ 
2002cr & $-$04.2(0.5) & May  10.1 & IRTF  &  0.0096 &  1.26$^{5}$      \\ 
2005am & $-$02.9(0.5) & Mar. 05.3 & IRTF  &  0.0079 &  0.78$^{2}$            \\ 
2011fe & $-$02.8(0.6) & Sep. 07.1 & Gem-N &  0.0008 &  1.07$^{8}$     \\
2002hw & $-$02.6(0.5) & Nov. 14.3 & IRTF  &  0.0175 &  1.51$^{5}$      \\ 
2002el & $-$01.8(0.5) & Aug. 20.5 & IRTF  &  0.0233 &  1.39$^{3}$    \\  
2001br & $-$00.9(0.6) & May  22.6 & IRTF  &  0.0206 &  0.93$^{3}$    \\ 
2004bl & $-$00.4(3.7) & May  08.3 & IRTF  &  0.0173 &  $\cdots$  \\ 
2005cf & $-$00.3(0.5) & Jun. 11.1 & VLT   &  0.0065 &  1.12$^{9}$   \\
2011fe & $+$00.3(0.5) & Sep. 10.2 & Gem-N &  0.0008 &  1.07$^{8}$     \\
1999ee & $+$00.5(0.5) & Oct. 18.0 & NTT   &  0.0114 &  0.94$^{7}$       \\ 
2005am & $+$01.2(0.5) & Mar. 09.4 & IRTF  &  0.0079 &  0.78$^{2}$            \\ 
1999ee & $+$01.5(0.5) & Oct. 19.0 & VLT   &  0.0114 &  0.94$^{7}$       \\ 
2005cf & $+$01.8(0.5) & Jun. 13.1 & VLT   &  0.0065 &  1.12$^{9}$    \\
2000dm & $+$02.3(0.6) & Oct. 01.2 & IRTF  &  0.0153 &  1.56$^{3}$    \\ 
2001dl & $+$02.5(0.5) & Aug. 12.5 & IRTF  &  0.0207 &  0.98$^{3}$    \\  
2003du & $+$02.7(0.5) & May  09.3 & IRTF  &  0.0168 &  1.02$^{6}$     \\  
2011fe & $+$03.3(0.5) & Sep. 13.2 & Gem-N &  0.0008 &  1.07$^{8}$     \\
2003du & $+$04.5(0.5) & May  11.0 & UKIRT &  0.0064 &  1.02$^{6}$     \\ 
1999ee & $+$04.5(0.5) & Oct. 22.1 & NTT   &  0.0114 &  0.94$^{7}$       \\ 
2001bf & $+$05.7(0.5) & May  21.5 & IRTF  &  0.0155 &  0.93$^{3}$    \\ 
2004da & $+$06.0(3.0) & Jul. 08.5 & IRTF  &  0.0163 &  0.87$^{4}$                  \\ 
2000do & $+$06.2(3.0) & Oct. 02.2 & IRTF  &  0.0109 &  $\cdots$        \\ 
2000dk & $+$06.3(0.5) & Oct. 01.5 & IRTF  &  0.0174 &  1.63$^{1}$             \\   
2005am & $+$08.2(0.5) & Mar. 16.4 & IRTF  &  0.0079 &  0.78$^{2}$            \\ 
2011fe & $+$08.2(0.5) & Sep. 18.3 & Gem-N &  0.0008 &  1.07$^{8}$     \\
1999ee & $+$08.5(0.5) & Oct. 26.0 & NTT   &  0.0114 &  0.94$^{7}$       \\ 
2005cf & $+$10.7(0.5) & Jun. 22.1 & VLT   &  0.0065 &  1.12$^{9}$     \\
2002ha & $+$11.3(0.5) & Nov. 14.2 & IRTF  &  0.0141 &  1.34$^{5}$      \\
2001bg & $+$11.3(0.5) & May  22.2 & IRTF  &  0.0071 &  1.10$^{3}$    \\  
2011fe & $+$12.3(0.5) & Sep. 22.2 & Gem-N &  0.0008 &  1.07$^{8}$     \\
2005cf & $+$14.4(0.5) & Jun. 27.1 & TNG   &  0.0065 &  1.12$^{9}$  \\
2004ab & $+$15.0(3.0) & Mar. 07.5 & IRTF  &  0.0058 &  $\cdots$         \\
2005am & $+$15.1(0.5) & Mar. 23.3 & IRTF  &  0.0079 &  0.78$^{2}$            \\
2003du & $+$15.4(0.5) & May  22.0 & TNG   &  0.0064 &  1.02$^{6}$     \\
1999ee & $+$15.5(0.5) & Nov. 02.0 & VLT   &  0.0114 &  0.94$^{7}$       \\ 
2002ef & $+$15.6(0.6) & Aug. 20.6 & IRTF  &  0.0240 &  1.04$^{3}$    \\ 
2004da & $+$15.9(3.0) & Jul. 18.4 & IRTF  &  0.0163 &  0.87$^{4}$             \\
2003du & $+$16.2(0.5) & May  23.1 & TNG   &  0.0064 &  1.02$^{6}$   \\
2011fe & $+$17.3(0.5) & Sep. 27.2 & Gem-N &  0.0008 &  1.07$^{8}$     \\
2004bk & $+$19.7(0.7) & May  08.5 & IRTF  &  0.0230 &  1.18$^{3}$    \\
2001en & $+$19.8(0.5) & Oct. 30.3 & IRTF  &  0.0159 &  1.27$^{3}$    \\ 
2004da & $+$20.0(3.0) & Jul. 22.5 & IRTF  &  0.0163 &  0.87$^{4}$             \\
2003du & $+$20.4(0.5) & May  27.0 & UKIRT &  0.0064 &  1.02$^{6}$     \\  
1999ee & $+$22.5(0.5) & Nov. 09.0 & NTT   &  0.0114 &  0.94$^{7}$       \\ 
2004da & $+$25.9(3.0) & Jul. 28.4 & IRTF  &  0.0163 &  0.87$^{4}$             \\  
1999ee & $+$27.6(0.5) & Nov. 14.1 & NTT   &  0.0114 &  0.94$^{7}$       \\  
2003du & $+$30.4(0.5) & Jun. 06.0 & UKIRT &  0.0064 &  1.02$^{6}$     \\  
2005cf & $+$31.4(0.5) & Jul. 14.1 & TNG   &  0.0065 &  1.12$^{9}$    \\
1999ee & $+$31.5(0.5) & Nov. 18.0 & VLT   &  0.0114 &  0.94$^{7}$       \\ 
2004E  & $+$38.2(2.5) & Feb. 21.5 & IRTF  &  0.0298 &  1.27$^{3}$    \\  
2003cg & $+$39.3(0.5) & May  09.2 & IRTF  &  0.0041 &  1.25$^{1}$             \\  
1999ee & $+$41.5(0.5) & Nov. 28.0 & VLT   &  0.0114 &  0.94$^{7}$       \\ 
2005cf & $+$41.5(0.5) & Jul. 24.0 & TNG   &  0.0065 &  1.12$^{9}$     \\
2002fk & $+$44.8(0.5) & Nov. 14.4 & IRTF  &  0.0071 &  1.08$^{3}$    \\   
2004ca & $+$45.1(3.0) & Jul. 28.6 & IRTF  &  0.0173 &  $\cdots$         \\  
2001gc & $+$48.3(5.5) & Jan. 14.5 & IRTF  &  0.0193 &  1.28$^{5}$      \\
2004bv & $+$53.8(0.5) & Jul. 28.3 & IRTF  &  0.0105 &  0.89$^{3}$    \\ 
\enddata
\tablenotetext{a}{Estimates of $T(B_{max})$ with a 3.0 day uncertainty were
determined from cross-correlation with a set of template spectra using SNID 
\citep{blondin07}.}
\tablenotetext{b}{Estimates of $\Delta m_{15}$  
are from the following references: 
(1)   \citet{eli06};
(2)    \citet{folatelli10};
(3) \citet{mo10};
(4)  Marion, private communication: McDonald 30in, unpublished $UBVRI$ photometry;
(5)  \citet{hicken09};
(6) \citet{stanishev07};
(7) \citet{stritzinger02};
(8) \citet{vinko12};
(9) \citet{pastorello07}.}
\end{deluxetable}

\clearpage
\begin{deluxetable}{rrcrcrcrc}
\tablewidth{0pt}
\tablenum{2}
\tablecaption{$YJHK_s$$-$band  $K$~corrections at $z=0.03, 0.05, 0.08$\label{tabkcor}}
\tablehead{
\colhead{Phase} &
\colhead{$Y$} &
\colhead{$\sigma_{stat}$} &
\colhead{$J$} &
\colhead{$\sigma_{stat}$} &
\colhead{$H$} &
\colhead{$\sigma_{stat}$} &
\colhead{$K_s$} &
\colhead{$\sigma_{stat}$}}
\startdata
\multicolumn{9}{c}{\bf $z=0.03$}\\
-14.6 & -0.075 & $\pm$0.011 & -0.061 & $\pm$0.016 & -0.050 & $\pm$0.016 & -0.024 & $\pm$0.017 \\ 
-13.9 & -0.072 & $\pm$0.009 & -0.059 & $\pm$0.014 & -0.049 & $\pm$0.013 & -0.026 & $\pm$0.015 \\
-13.2 & -0.070 & $\pm$0.008 & -0.057 & $\pm$0.012 & -0.048 & $\pm$0.011 & -0.028 & $\pm$0.013 \\
-12.5 & -0.068 & $\pm$0.007 & -0.055 & $\pm$0.011 & -0.046 & $\pm$0.010 & -0.030 & $\pm$0.011 \\
-11.8 & -0.066 & $\pm$0.006 & -0.053 & $\pm$0.010 & -0.044 & $\pm$0.008 & -0.033 & $\pm$0.010 \\
-11.1 & -0.065 & $\pm$0.005 & -0.051 & $\pm$0.009 & -0.042 & $\pm$0.008 & -0.036 & $\pm$0.008 \\
-10.5 & -0.065 & $\pm$0.005 & -0.048 & $\pm$0.008 & -0.040 & $\pm$0.007 & -0.039 & $\pm$0.008 \\
-9.8  & -0.065 & $\pm$0.005 & -0.046 & $\pm$0.008 & -0.038 & $\pm$0.007 & -0.043 & $\pm$0.008 \\ 
-9.1  & -0.065 & $\pm$0.005 & -0.044 & $\pm$0.008 & -0.037 & $\pm$0.007 & -0.047 & $\pm$0.007 \\ 
-8.4  & -0.067 & $\pm$0.004 & -0.041 & $\pm$0.007 & -0.035 & $\pm$0.007 & -0.051 & $\pm$0.007 \\ 
-7.7  & -0.068 & $\pm$0.004 & -0.039 & $\pm$0.007 & -0.033 & $\pm$0.006 & -0.056 & $\pm$0.007 \\ 
-7.0  & -0.071 & $\pm$0.004 & -0.037 & $\pm$0.007 & -0.032 & $\pm$0.006 & -0.061 & $\pm$0.007 \\ 
-6.3  & -0.074 & $\pm$0.004 & -0.034 & $\pm$0.007 & -0.031 & $\pm$0.006 & -0.066 & $\pm$0.007 \\ 
-5.6  & -0.077 & $\pm$0.004 & -0.032 & $\pm$0.007 & -0.030 & $\pm$0.006 & -0.072 & $\pm$0.007 \\ 
-4.9  & -0.080 & $\pm$0.004 & -0.030 & $\pm$0.007 & -0.030 & $\pm$0.006 & -0.078 & $\pm$0.007 \\ 
-4.2  & -0.084 & $\pm$0.004 & -0.027 & $\pm$0.007 & -0.029 & $\pm$0.006 & -0.084 & $\pm$0.007 \\ 
-3.5  & -0.088 & $\pm$0.004 & -0.024 & $\pm$0.007 & -0.028 & $\pm$0.006 & -0.090 & $\pm$0.007 \\ 
-2.9  & -0.092 & $\pm$0.004 & -0.022 & $\pm$0.007 & -0.028 & $\pm$0.006 & -0.096 & $\pm$0.007 \\ 
-2.2  & -0.096 & $\pm$0.004 & -0.019 & $\pm$0.007 & -0.026 & $\pm$0.006 & -0.103 & $\pm$0.007 \\ 
-1.5  & -0.101 & $\pm$0.004 & -0.016 & $\pm$0.006 & -0.024 & $\pm$0.006 & -0.109 & $\pm$0.007 \\ 
-0.8  & -0.105 & $\pm$0.004 & -0.014 & $\pm$0.006 & -0.022 & $\pm$0.006 & -0.115 & $\pm$0.007 \\ 
-0.1  & -0.109 & $\pm$0.004 & -0.012 & $\pm$0.007 & -0.018 & $\pm$0.006 & -0.121 & $\pm$0.007 \\ 
0.6   & -0.113 & $\pm$0.004 & -0.011 & $\pm$0.007 & -0.013 & $\pm$0.006 & -0.127 & $\pm$0.007 \\  
1.3   & -0.117 & $\pm$0.004 & -0.010 & $\pm$0.007 & -0.007 & $\pm$0.006 & -0.132 & $\pm$0.007 \\  
2.0   & -0.122 & $\pm$0.004 & -0.009 & $\pm$0.007 &  0.000 & $\pm$0.006 & -0.137 & $\pm$0.007 \\  
2.7   & -0.126 & $\pm$0.004 & -0.010 & $\pm$0.007 &  0.009 & $\pm$0.006 & -0.142 & $\pm$0.007 \\  
3.4   & -0.130 & $\pm$0.004 & -0.010 & $\pm$0.007 &  0.019 & $\pm$0.006 & -0.146 & $\pm$0.007 \\  
4.1   & -0.134 & $\pm$0.004 & -0.012 & $\pm$0.008 &  0.031 & $\pm$0.006 & -0.149 & $\pm$0.007 \\  
4.7   & -0.139 & $\pm$0.004 & -0.014 & $\pm$0.008 &  0.043 & $\pm$0.006 & -0.151 & $\pm$0.007 \\  
5.4   & -0.143 & $\pm$0.004 & -0.016 & $\pm$0.008 &  0.057 & $\pm$0.006 & -0.153 & $\pm$0.008 \\  
6.1   & -0.147 & $\pm$0.005 & -0.019 & $\pm$0.008 &  0.072 & $\pm$0.006 & -0.154 & $\pm$0.008 \\  
6.8   & -0.151 & $\pm$0.005 & -0.023 & $\pm$0.008 &  0.087 & $\pm$0.007 & -0.155 & $\pm$0.008 \\  
7.5   & -0.154 & $\pm$0.005 & -0.027 & $\pm$0.008 &  0.101 & $\pm$0.007 & -0.154 & $\pm$0.008 \\  
8.2   & -0.158 & $\pm$0.005 & -0.031 & $\pm$0.008 &  0.116 & $\pm$0.008 & -0.153 & $\pm$0.008 \\  
8.9   & -0.162 & $\pm$0.005 & -0.036 & $\pm$0.008 &  0.129 & $\pm$0.008 & -0.152 & $\pm$0.008 \\  
9.6   & -0.165 & $\pm$0.005 & -0.042 & $\pm$0.007 &  0.141 & $\pm$0.009 & -0.149 & $\pm$0.008 \\  
10.3  & -0.168 & $\pm$0.005 & -0.048 & $\pm$0.007 &  0.151 & $\pm$0.009 & -0.146 & $\pm$0.009 \\ 
11.0  & -0.171 & $\pm$0.005 & -0.054 & $\pm$0.007 &  0.159 & $\pm$0.009 & -0.142 & $\pm$0.008 \\
11.7  & -0.173 & $\pm$0.005 & -0.061 & $\pm$0.008 &  0.165 & $\pm$0.009 & -0.137 & $\pm$0.008 \\
12.3  & -0.175 & $\pm$0.005 & -0.067 & $\pm$0.008 &  0.168 & $\pm$0.009 & -0.132 & $\pm$0.008 \\
13.0  & -0.177 & $\pm$0.005 & -0.074 & $\pm$0.009 &  0.170 & $\pm$0.009 & -0.126 & $\pm$0.008 \\
13.7  & -0.178 & $\pm$0.005 & -0.080 & $\pm$0.009 &  0.169 & $\pm$0.008 & -0.120 & $\pm$0.008 \\
14.4  & -0.179 & $\pm$0.005 & -0.086 & $\pm$0.009 &  0.168 & $\pm$0.008 & -0.114 & $\pm$0.008 \\ 
15.1  & -0.179 & $\pm$0.005 & -0.092 & $\pm$0.009 &  0.165 & $\pm$0.008 & -0.107 & $\pm$0.008 \\ 
15.8  & -0.178 & $\pm$0.005 & -0.097 & $\pm$0.009 &  0.161 & $\pm$0.008 & -0.100 & $\pm$0.008 \\ 
16.5  & -0.176 & $\pm$0.005 & -0.103 & $\pm$0.008 &  0.157 & $\pm$0.008 & -0.092 & $\pm$0.008 \\ 
17.2  & -0.173 & $\pm$0.005 & -0.108 & $\pm$0.008 &  0.153 & $\pm$0.008 & -0.085 & $\pm$0.008 \\ 
17.9  & -0.170 & $\pm$0.005 & -0.113 & $\pm$0.008 &  0.148 & $\pm$0.008 & -0.078 & $\pm$0.008 \\ 
18.6  & -0.165 & $\pm$0.006 & -0.118 & $\pm$0.008 &  0.144 & $\pm$0.008 & -0.070 & $\pm$0.009 \\ 
19.2  & -0.160 & $\pm$0.006 & -0.122 & $\pm$0.008 &  0.139 & $\pm$0.008 & -0.063 & $\pm$0.009 \\ 
19.9  & -0.154 & $\pm$0.006 & -0.127 & $\pm$0.009 &  0.135 & $\pm$0.009 & -0.056 & $\pm$0.010 \\ 
20.6  & -0.147 & $\pm$0.006 & -0.131 & $\pm$0.009 &  0.131 & $\pm$0.009 & -0.049 & $\pm$0.010 \\ 
21.3  & -0.140 & $\pm$0.006 & -0.136 & $\pm$0.010 &  0.128 & $\pm$0.009 & -0.042 & $\pm$0.010 \\ 
22.0  & -0.132 & $\pm$0.007 & -0.140 & $\pm$0.011 &  0.125 & $\pm$0.010 & -0.036 & $\pm$0.010 \\ 
22.7  & -0.123 & $\pm$0.007 & -0.145 & $\pm$0.012 &  0.122 & $\pm$0.010 & -0.030 & $\pm$0.011 \\ 
23.4  & -0.114 & $\pm$0.007 & -0.149 & $\pm$0.012 &  0.119 & $\pm$0.011 & -0.024 & $\pm$0.011 \\ 
24.1  & -0.105 & $\pm$0.007 & -0.154 & $\pm$0.013 &  0.116 & $\pm$0.011 & -0.018 & $\pm$0.012 \\ 
24.8  & -0.095 & $\pm$0.007 & -0.159 & $\pm$0.013 &  0.113 & $\pm$0.011 & -0.013 & $\pm$0.012 \\ 
25.5  & -0.086 & $\pm$0.007 & -0.164 & $\pm$0.013 &  0.110 & $\pm$0.011 & -0.008 & $\pm$0.012 \\ 
26.2  & -0.077 & $\pm$0.007 & -0.169 & $\pm$0.013 &  0.107 & $\pm$0.011 & -0.003 & $\pm$0.012 \\ 
26.8  & -0.068 & $\pm$0.007 & -0.174 & $\pm$0.012 &  0.103 & $\pm$0.011 &  0.001 & $\pm$0.012 \\ 
27.5  & -0.059 & $\pm$0.007 & -0.179 & $\pm$0.012 &  0.099 & $\pm$0.011 &  0.005 & $\pm$0.013 \\ 
28.2  & -0.050 & $\pm$0.007 & -0.185 & $\pm$0.012 &  0.095 & $\pm$0.011 &  0.008 & $\pm$0.013 \\ 
28.9  & -0.042 & $\pm$0.007 & -0.190 & $\pm$0.012 &  0.090 & $\pm$0.011 &  0.012 & $\pm$0.013 \\ 
29.6  & -0.034 & $\pm$0.007 & -0.195 & $\pm$0.012 &  0.085 & $\pm$0.011 &  0.014 & $\pm$0.013 \\ 
30.3  & -0.027 & $\pm$0.008 & -0.200 & $\pm$0.012 &  0.079 & $\pm$0.011 &  0.017 & $\pm$0.013 \\ 
31.0  & -0.020 & $\pm$0.008 & -0.204 & $\pm$0.012 &  0.074 & $\pm$0.011 &  0.019 & $\pm$0.013 \\ 
31.7  & -0.014 & $\pm$0.008 & -0.209 & $\pm$0.012 &  0.069 & $\pm$0.011 &  0.020 & $\pm$0.013 \\ 
32.4  & -0.009 & $\pm$0.008 & -0.213 & $\pm$0.013 &  0.064 & $\pm$0.012 &  0.021 & $\pm$0.014 \\ 
33.1  & -0.004 & $\pm$0.009 & -0.216 & $\pm$0.014 &  0.059 & $\pm$0.012 &  0.022 & $\pm$0.014 \\ 
33.8  & -0.000 & $\pm$0.009 & -0.220 & $\pm$0.015 &  0.055 & $\pm$0.013 &  0.023 & $\pm$0.014 \\ 
34.4  &  0.003 & $\pm$0.009 & -0.223 & $\pm$0.016 &  0.052 & $\pm$0.014 &  0.023 & $\pm$0.014 \\ 
35.1  &  0.006 & $\pm$0.009 & -0.227 & $\pm$0.016 &  0.049 & $\pm$0.014 &  0.023 & $\pm$0.014 \\ 
35.8  &  0.008 & $\pm$0.009 & -0.230 & $\pm$0.016 &  0.046 & $\pm$0.014 &  0.022 & $\pm$0.014 \\ 
36.5  &  0.010 & $\pm$0.009 & -0.232 & $\pm$0.016 &  0.044 & $\pm$0.014 &  0.022 & $\pm$0.014 \\ 
37.2  &  0.012 & $\pm$0.009 & -0.235 & $\pm$0.016 &  0.044 & $\pm$0.013 &  0.021 & $\pm$0.014 \\ 
37.9  &  0.013 & $\pm$0.009 & -0.238 & $\pm$0.015 &  0.043 & $\pm$0.012 &  0.020 & $\pm$0.014 \\ 
38.6  &  0.014 & $\pm$0.008 & -0.240 & $\pm$0.014 &  0.044 & $\pm$0.012 &  0.019 & $\pm$0.013 \\ 
39.3  &  0.015 & $\pm$0.008 & -0.242 & $\pm$0.013 &  0.045 & $\pm$0.011 &  0.017 & $\pm$0.013 \\ 
40.0  &  0.016 & $\pm$0.008 & -0.244 & $\pm$0.013 &  0.047 & $\pm$0.010 &  0.016 & $\pm$0.013 \\ 
40.7  &  0.016 & $\pm$0.008 & -0.246 & $\pm$0.013 &  0.050 & $\pm$0.010 &  0.015 & $\pm$0.013 \\ 
41.4  &  0.017 & $\pm$0.008 & -0.247 & $\pm$0.014 &  0.053 & $\pm$0.010 &  0.013 & $\pm$0.012 \\ 
42.0  &  0.017 & $\pm$0.008 & -0.248 & $\pm$0.015 &  0.056 & $\pm$0.010 &  0.012 & $\pm$0.012 \\ 
42.7  &  0.017 & $\pm$0.008 & -0.248 & $\pm$0.015 &  0.060 & $\pm$0.010 &  0.011 & $\pm$0.013 \\ 
43.4  &  0.017 & $\pm$0.008 & -0.247 & $\pm$0.015 &  0.064 & $\pm$0.010 &  0.009 & $\pm$0.013 \\ 
44.1  &  0.018 & $\pm$0.008 & -0.246 & $\pm$0.014 &  0.068 & $\pm$0.010 &  0.008 & $\pm$0.013 \\ 
44.8  &  0.018 & $\pm$0.008 & -0.245 & $\pm$0.013 &  0.071 & $\pm$0.010 &  0.007 & $\pm$0.013 \\ 
45.5  &  0.018 & $\pm$0.008 & -0.243 & $\pm$0.013 &  0.075 & $\pm$0.011 &  0.006 & $\pm$0.014 \\ 
46.2  &  0.019 & $\pm$0.009 & -0.241 & $\pm$0.015 &  0.078 & $\pm$0.011 &  0.005 & $\pm$0.014 \\ 
46.9  &  0.019 & $\pm$0.009 & -0.239 & $\pm$0.019 &  0.081 & $\pm$0.012 &  0.004 & $\pm$0.014 \\ 
47.6  &  0.020 & $\pm$0.009 & -0.237 & $\pm$0.022 &  0.084 & $\pm$0.012 &  0.003 & $\pm$0.015 \\ 
48.3  &  0.021 & $\pm$0.010 & -0.236 & $\pm$0.025 &  0.086 & $\pm$0.013 &  0.003 & $\pm$0.015 \\ 
49.0  &  0.022 & $\pm$0.010 & -0.236 & $\pm$0.027 &  0.089 & $\pm$0.014 &  0.002 & $\pm$0.016 \\ 
49.6  &  0.022 & $\pm$0.010 & -0.236 & $\pm$0.028 &  0.091 & $\pm$0.014 &  0.002 & $\pm$0.016 \\ 
50.3  &  0.023 & $\pm$0.010 & -0.237 & $\pm$0.027 &  0.093 & $\pm$0.015 &  0.001 & $\pm$0.017 \\ 
51.0  &  0.024 & $\pm$0.011 & -0.240 & $\pm$0.025 &  0.095 & $\pm$0.015 &  0.001 & $\pm$0.018 \\ 
51.7  &  0.025 & $\pm$0.011 & -0.242 & $\pm$0.023 &  0.097 & $\pm$0.015 &  0.001 & $\pm$0.019 \\ 
52.4  &  0.025 & $\pm$0.012 & -0.246 & $\pm$0.021 &  0.099 & $\pm$0.016 &  0.001 & $\pm$0.020 \\ 
53.1  &  0.026 & $\pm$0.013 & -0.250 & $\pm$0.022 &  0.100 & $\pm$0.017 &  0.000 & $\pm$0.022 \\ 
53.8  &  0.027 & $\pm$0.014 & -0.254 & $\pm$0.026 &  0.102 & $\pm$0.018 &  0.000 & $\pm$0.023 \\ 
\multicolumn{9}{c}{\bf $z=0.05$}\\
-14.6 & -0.144 & $\pm$0.018 & -0.088 & $\pm$0.026 & -0.063 & $\pm$0.038 & -0.040 & $\pm$0.028 \\           
-13.9 & -0.138 & $\pm$0.015 & -0.085 & $\pm$0.022 & -0.064 & $\pm$0.032 & -0.042 & $\pm$0.024 \\
-13.2 & -0.133 & $\pm$0.012 & -0.082 & $\pm$0.018 & -0.065 & $\pm$0.028 & -0.044 & $\pm$0.021 \\
-12.5 & -0.128 & $\pm$0.010 & -0.078 & $\pm$0.016 & -0.065 & $\pm$0.024 & -0.048 & $\pm$0.018 \\
-11.8 & -0.124 & $\pm$0.009 & -0.075 & $\pm$0.014 & -0.066 & $\pm$0.021 & -0.051 & $\pm$0.016 \\
-11.1 & -0.121 & $\pm$0.008 & -0.071 & $\pm$0.013 & -0.066 & $\pm$0.019 & -0.056 & $\pm$0.015 \\
-10.5 & -0.118 & $\pm$0.008 & -0.068 & $\pm$0.012 & -0.066 & $\pm$0.018 & -0.061 & $\pm$0.013 \\
-9.8  & -0.117 & $\pm$0.007 & -0.064 & $\pm$0.012 & -0.066 & $\pm$0.017 & -0.067 & $\pm$0.013 \\
-9.1  & -0.116 & $\pm$0.007 & -0.061 & $\pm$0.012 & -0.066 & $\pm$0.016 & -0.073 & $\pm$0.012 \\
-8.4  & -0.117 & $\pm$0.007 & -0.057 & $\pm$0.011 & -0.067 & $\pm$0.016 & -0.080 & $\pm$0.012 \\
-7.7  & -0.118 & $\pm$0.007 & -0.053 & $\pm$0.011 & -0.069 & $\pm$0.016 & -0.088 & $\pm$0.011 \\
-7.0  & -0.120 & $\pm$0.007 & -0.050 & $\pm$0.011 & -0.071 & $\pm$0.015 & -0.096 & $\pm$0.011 \\
-6.3  & -0.124 & $\pm$0.007 & -0.046 & $\pm$0.011 & -0.074 & $\pm$0.015 & -0.105 & $\pm$0.011 \\
-5.6  & -0.128 & $\pm$0.007 & -0.042 & $\pm$0.011 & -0.077 & $\pm$0.014 & -0.115 & $\pm$0.011 \\
-4.9  & -0.133 & $\pm$0.006 & -0.037 & $\pm$0.012 & -0.081 & $\pm$0.014 & -0.125 & $\pm$0.011 \\
-4.2  & -0.139 & $\pm$0.006 & -0.032 & $\pm$0.012 & -0.085 & $\pm$0.014 & -0.135 & $\pm$0.011 \\
-3.5  & -0.145 & $\pm$0.006 & -0.027 & $\pm$0.012 & -0.088 & $\pm$0.014 & -0.146 & $\pm$0.011 \\
-2.9  & -0.152 & $\pm$0.006 & -0.021 & $\pm$0.012 & -0.090 & $\pm$0.015 & -0.157 & $\pm$0.011 \\
-2.2  & -0.159 & $\pm$0.006 & -0.015 & $\pm$0.012 & -0.091 & $\pm$0.015 & -0.168 & $\pm$0.011 \\
-1.5  & -0.166 & $\pm$0.006 & -0.009 & $\pm$0.011 & -0.089 & $\pm$0.015 & -0.180 & $\pm$0.011 \\
-0.8  & -0.174 & $\pm$0.006 & -0.003 & $\pm$0.011 & -0.086 & $\pm$0.016 & -0.191 & $\pm$0.011 \\
-0.1  & -0.182 & $\pm$0.006 &  0.002 & $\pm$0.011 & -0.081 & $\pm$0.016 & -0.201 & $\pm$0.011 \\
0.6   & -0.189 & $\pm$0.006 &  0.006 & $\pm$0.011 & -0.073 & $\pm$0.016 & -0.211 & $\pm$0.011 \\
1.3   & -0.197 & $\pm$0.006 &  0.009 & $\pm$0.011 & -0.063 & $\pm$0.016 & -0.220 & $\pm$0.011 \\
2.0   & -0.204 & $\pm$0.006 &  0.011 & $\pm$0.011 & -0.050 & $\pm$0.015 & -0.229 & $\pm$0.011 \\
2.7   & -0.212 & $\pm$0.006 &  0.011 & $\pm$0.011 & -0.035 & $\pm$0.015 & -0.236 & $\pm$0.011 \\
3.4   & -0.219 & $\pm$0.006 &  0.009 & $\pm$0.012 & -0.019 & $\pm$0.015 & -0.242 & $\pm$0.012 \\
4.1   & -0.226 & $\pm$0.006 &  0.006 & $\pm$0.012 & -0.001 & $\pm$0.015 & -0.246 & $\pm$0.012 \\
4.7   & -0.233 & $\pm$0.007 &  0.001 & $\pm$0.012 &  0.019 & $\pm$0.015 & -0.249 & $\pm$0.012 \\
5.4   & -0.239 & $\pm$0.007 & -0.005 & $\pm$0.012 &  0.040 & $\pm$0.015 & -0.251 & $\pm$0.013 \\
6.1   & -0.246 & $\pm$0.007 & -0.012 & $\pm$0.012 &  0.061 & $\pm$0.016 & -0.251 & $\pm$0.013 \\
6.8   & -0.252 & $\pm$0.007 & -0.020 & $\pm$0.013 &  0.082 & $\pm$0.017 & -0.250 & $\pm$0.013 \\
7.5   & -0.258 & $\pm$0.008 & -0.030 & $\pm$0.013 &  0.103 & $\pm$0.018 & -0.247 & $\pm$0.013 \\
8.2   & -0.264 & $\pm$0.008 & -0.040 & $\pm$0.013 &  0.122 & $\pm$0.019 & -0.243 & $\pm$0.013 \\
8.9   & -0.270 & $\pm$0.008 & -0.052 & $\pm$0.013 &  0.141 & $\pm$0.021 & -0.238 & $\pm$0.013 \\
9.6   & -0.275 & $\pm$0.009 & -0.065 & $\pm$0.013 &  0.157 & $\pm$0.021 & -0.232 & $\pm$0.013 \\
10.3  & -0.281 & $\pm$0.009 & -0.079 & $\pm$0.013 &  0.172 & $\pm$0.022 & -0.225 & $\pm$0.013 \\
11.0  & -0.286 & $\pm$0.009 & -0.093 & $\pm$0.013 &  0.184 & $\pm$0.022 & -0.217 & $\pm$0.013 \\
11.7  & -0.291 & $\pm$0.009 & -0.108 & $\pm$0.013 &  0.194 & $\pm$0.021 & -0.208 & $\pm$0.013 \\
12.3  & -0.296 & $\pm$0.009 & -0.123 & $\pm$0.014 &  0.202 & $\pm$0.021 & -0.199 & $\pm$0.013 \\
13.0  & -0.300 & $\pm$0.009 & -0.137 & $\pm$0.015 &  0.207 & $\pm$0.020 & -0.189 & $\pm$0.012 \\
13.7  & -0.303 & $\pm$0.009 & -0.150 & $\pm$0.015 &  0.211 & $\pm$0.019 & -0.178 & $\pm$0.012 \\
14.4  & -0.306 & $\pm$0.009 & -0.163 & $\pm$0.014 &  0.213 & $\pm$0.018 & -0.167 & $\pm$0.012 \\
15.1  & -0.307 & $\pm$0.009 & -0.174 & $\pm$0.014 &  0.213 & $\pm$0.018 & -0.156 & $\pm$0.012 \\
15.8  & -0.306 & $\pm$0.009 & -0.185 & $\pm$0.013 &  0.213 & $\pm$0.018 & -0.144 & $\pm$0.012 \\
16.5  & -0.305 & $\pm$0.009 & -0.195 & $\pm$0.012 &  0.211 & $\pm$0.018 & -0.132 & $\pm$0.012 \\
17.2  & -0.301 & $\pm$0.009 & -0.205 & $\pm$0.013 &  0.209 & $\pm$0.018 & -0.120 & $\pm$0.012 \\
17.9  & -0.296 & $\pm$0.009 & -0.214 & $\pm$0.013 &  0.207 & $\pm$0.019 & -0.109 & $\pm$0.013 \\
18.6  & -0.290 & $\pm$0.010 & -0.222 & $\pm$0.013 &  0.205 & $\pm$0.020 & -0.097 & $\pm$0.013 \\
19.2  & -0.281 & $\pm$0.010 & -0.231 & $\pm$0.014 &  0.202 & $\pm$0.020 & -0.085 & $\pm$0.014 \\
19.9  & -0.271 & $\pm$0.010 & -0.239 & $\pm$0.015 &  0.199 & $\pm$0.021 & -0.074 & $\pm$0.015 \\
20.6  & -0.260 & $\pm$0.010 & -0.247 & $\pm$0.016 &  0.197 & $\pm$0.022 & -0.063 & $\pm$0.016 \\
21.3  & -0.247 & $\pm$0.010 & -0.256 & $\pm$0.017 &  0.194 & $\pm$0.022 & -0.052 & $\pm$0.017 \\
22.0  & -0.232 & $\pm$0.011 & -0.264 & $\pm$0.019 &  0.191 & $\pm$0.023 & -0.042 & $\pm$0.017 \\
22.7  & -0.217 & $\pm$0.011 & -0.272 & $\pm$0.021 &  0.189 & $\pm$0.024 & -0.032 & $\pm$0.018 \\
23.4  & -0.201 & $\pm$0.011 & -0.281 & $\pm$0.022 &  0.186 & $\pm$0.024 & -0.022 & $\pm$0.019 \\
24.1  & -0.184 & $\pm$0.011 & -0.290 & $\pm$0.022 &  0.183 & $\pm$0.025 & -0.013 & $\pm$0.019 \\
24.8  & -0.166 & $\pm$0.012 & -0.300 & $\pm$0.023 &  0.181 & $\pm$0.026 & -0.004 & $\pm$0.019 \\
25.5  & -0.149 & $\pm$0.012 & -0.310 & $\pm$0.022 &  0.178 & $\pm$0.026 &  0.004 & $\pm$0.020 \\
26.2  & -0.131 & $\pm$0.012 & -0.321 & $\pm$0.022 &  0.176 & $\pm$0.026 &  0.012 & $\pm$0.020 \\
26.8  & -0.113 & $\pm$0.012 & -0.332 & $\pm$0.021 &  0.173 & $\pm$0.026 &  0.019 & $\pm$0.020 \\
27.5  & -0.095 & $\pm$0.012 & -0.343 & $\pm$0.020 &  0.171 & $\pm$0.026 &  0.025 & $\pm$0.020 \\
28.2  & -0.078 & $\pm$0.012 & -0.354 & $\pm$0.020 &  0.168 & $\pm$0.026 &  0.031 & $\pm$0.020 \\
28.9  & -0.062 & $\pm$0.012 & -0.365 & $\pm$0.020 &  0.165 & $\pm$0.026 &  0.036 & $\pm$0.020 \\
29.6  & -0.046 & $\pm$0.012 & -0.375 & $\pm$0.021 &  0.161 & $\pm$0.026 &  0.041 & $\pm$0.020 \\
30.3  & -0.032 & $\pm$0.012 & -0.384 & $\pm$0.021 &  0.157 & $\pm$0.026 &  0.044 & $\pm$0.020 \\
31.0  & -0.018 & $\pm$0.012 & -0.391 & $\pm$0.021 &  0.153 & $\pm$0.026 &  0.047 & $\pm$0.020 \\
31.7  & -0.006 & $\pm$0.013 & -0.398 & $\pm$0.021 &  0.148 & $\pm$0.027 &  0.050 & $\pm$0.021 \\
32.4  &  0.005 & $\pm$0.014 & -0.403 & $\pm$0.022 &  0.143 & $\pm$0.028 &  0.051 & $\pm$0.021 \\
33.1  &  0.015 & $\pm$0.014 & -0.408 & $\pm$0.024 &  0.138 & $\pm$0.029 &  0.052 & $\pm$0.021 \\
33.8  &  0.024 & $\pm$0.015 & -0.411 & $\pm$0.026 &  0.134 & $\pm$0.029 &  0.052 & $\pm$0.022 \\
34.4  &  0.031 & $\pm$0.015 & -0.414 & $\pm$0.029 &  0.129 & $\pm$0.030 &  0.052 & $\pm$0.022 \\
35.1  &  0.038 & $\pm$0.016 & -0.417 & $\pm$0.031 &  0.125 & $\pm$0.031 &  0.051 & $\pm$0.022 \\
35.8  &  0.043 & $\pm$0.016 & -0.420 & $\pm$0.032 &  0.122 & $\pm$0.031 &  0.050 & $\pm$0.022 \\
36.5  &  0.048 & $\pm$0.016 & -0.423 & $\pm$0.032 &  0.119 & $\pm$0.031 &  0.049 & $\pm$0.022 \\
37.2  &  0.052 & $\pm$0.015 & -0.427 & $\pm$0.031 &  0.117 & $\pm$0.030 &  0.048 & $\pm$0.022 \\
37.9  &  0.056 & $\pm$0.015 & -0.430 & $\pm$0.029 &  0.116 & $\pm$0.029 &  0.046 & $\pm$0.022 \\
38.6  &  0.058 & $\pm$0.014 & -0.434 & $\pm$0.025 &  0.116 & $\pm$0.027 &  0.045 & $\pm$0.021 \\
39.3  &  0.060 & $\pm$0.014 & -0.438 & $\pm$0.022 &  0.117 & $\pm$0.026 &  0.043 & $\pm$0.021 \\
40.0  &  0.062 & $\pm$0.013 & -0.441 & $\pm$0.021 &  0.118 & $\pm$0.025 &  0.041 & $\pm$0.020 \\
40.7  &  0.063 & $\pm$0.013 & -0.444 & $\pm$0.022 &  0.120 & $\pm$0.024 &  0.040 & $\pm$0.020 \\
41.4  &  0.064 & $\pm$0.013 & -0.445 & $\pm$0.024 &  0.122 & $\pm$0.024 &  0.038 & $\pm$0.020 \\
42.0  &  0.064 & $\pm$0.013 & -0.445 & $\pm$0.026 &  0.125 & $\pm$0.024 &  0.037 & $\pm$0.020 \\
42.7  &  0.064 & $\pm$0.013 & -0.443 & $\pm$0.027 &  0.128 & $\pm$0.024 &  0.035 & $\pm$0.020 \\
43.4  &  0.064 & $\pm$0.013 & -0.438 & $\pm$0.027 &  0.130 & $\pm$0.024 &  0.034 & $\pm$0.020 \\
44.1  &  0.063 & $\pm$0.013 & -0.432 & $\pm$0.025 &  0.133 & $\pm$0.025 &  0.033 & $\pm$0.020 \\
44.8  &  0.063 & $\pm$0.014 & -0.424 & $\pm$0.023 &  0.136 & $\pm$0.026 &  0.032 & $\pm$0.020 \\
45.5  &  0.063 & $\pm$0.014 & -0.415 & $\pm$0.022 &  0.138 & $\pm$0.027 &  0.031 & $\pm$0.021 \\
46.2  &  0.063 & $\pm$0.015 & -0.406 & $\pm$0.026 &  0.141 & $\pm$0.028 &  0.031 & $\pm$0.021 \\
46.9  &  0.064 & $\pm$0.015 & -0.396 & $\pm$0.032 &  0.143 & $\pm$0.030 &  0.031 & $\pm$0.022 \\
47.6  &  0.065 & $\pm$0.016 & -0.388 & $\pm$0.039 &  0.145 & $\pm$0.032 &  0.030 & $\pm$0.023 \\
48.3  &  0.067 & $\pm$0.017 & -0.381 & $\pm$0.046 &  0.148 & $\pm$0.033 &  0.030 & $\pm$0.024 \\
49.0  &  0.070 & $\pm$0.017 & -0.376 & $\pm$0.050 &  0.150 & $\pm$0.035 &  0.030 & $\pm$0.025 \\
49.6  &  0.073 & $\pm$0.018 & -0.375 & $\pm$0.051 &  0.153 & $\pm$0.037 &  0.031 & $\pm$0.026 \\
50.3  &  0.077 & $\pm$0.018 & -0.375 & $\pm$0.049 &  0.157 & $\pm$0.039 &  0.031 & $\pm$0.027 \\
51.0  &  0.081 & $\pm$0.019 & -0.379 & $\pm$0.045 &  0.160 & $\pm$0.040 &  0.031 & $\pm$0.029 \\
51.7  &  0.085 & $\pm$0.020 & -0.385 & $\pm$0.040 &  0.163 & $\pm$0.042 &  0.032 & $\pm$0.031 \\
52.4  &  0.090 & $\pm$0.021 & -0.393 & $\pm$0.035 &  0.166 & $\pm$0.044 &  0.032 & $\pm$0.033 \\
53.1  &  0.095 & $\pm$0.022 & -0.402 & $\pm$0.034 &  0.169 & $\pm$0.047 &  0.032 & $\pm$0.035 \\
53.8  &  0.100 & $\pm$0.025 & -0.413 & $\pm$0.039 &  0.172 & $\pm$0.050 &  0.033 & $\pm$0.038 \\
\multicolumn{9}{c}{\bf $z=0.08$}\\ 
-14.6 & -0.174 &$\pm$0.028  & -0.118 & $\pm$0.048 & -0.050 & $\pm$0.057 & -0.092 & $\pm$0.039 \\    
-13.9 & -0.173 &$\pm$0.023  & -0.108 & $\pm$0.039 & -0.060 & $\pm$0.049 & -0.093 & $\pm$0.034 \\ 
-13.2 & -0.172 &$\pm$0.020  & -0.099 & $\pm$0.033 & -0.070 & $\pm$0.043 & -0.095 & $\pm$0.030 \\ 
-12.5 & -0.172 &$\pm$0.017  & -0.090 & $\pm$0.028 & -0.070 & $\pm$0.038 & -0.098 & $\pm$0.026 \\ 
-11.8 & -0.171 &$\pm$0.015  & -0.081 & $\pm$0.025 & -0.080 & $\pm$0.034 & -0.101 & $\pm$0.023 \\ 
-11.1 & -0.171 &$\pm$0.014  & -0.074 & $\pm$0.022 & -0.090 & $\pm$0.031 & -0.106 & $\pm$0.021 \\ 
-10.5 & -0.172 &$\pm$0.013  & -0.067 & $\pm$0.021 & -0.100 & $\pm$0.029 & -0.111 & $\pm$0.019 \\ 
-9.8  & -0.172 &$\pm$0.012  & -0.061 & $\pm$0.021 & -0.107 & $\pm$0.028 & -0.118 & $\pm$0.018 \\
-9.1  & -0.174 &$\pm$0.012  & -0.055 & $\pm$0.020 & -0.115 & $\pm$0.027 & -0.125 & $\pm$0.018 \\
-8.4  & -0.176 &$\pm$0.011  & -0.050 & $\pm$0.020 & -0.124 & $\pm$0.027 & -0.134 & $\pm$0.017 \\
-7.7  & -0.178 &$\pm$0.011  & -0.046 & $\pm$0.019 & -0.133 & $\pm$0.026 & -0.144 & $\pm$0.017 \\
-7.0  & -0.182 &$\pm$0.010  & -0.041 & $\pm$0.019 & -0.141 & $\pm$0.026 & -0.155 & $\pm$0.017 \\
-6.3  & -0.187 &$\pm$0.010  & -0.036 & $\pm$0.018 & -0.148 & $\pm$0.026 & -0.167 & $\pm$0.016 \\
-5.6  & -0.193 &$\pm$0.010  & -0.031 & $\pm$0.018 & -0.154 & $\pm$0.025 & -0.179 & $\pm$0.016 \\
-4.9  & -0.200 &$\pm$0.010  & -0.024 & $\pm$0.019 & -0.160 & $\pm$0.025 & -0.193 & $\pm$0.016 \\
-4.2  & -0.208 &$\pm$0.010  & -0.016 & $\pm$0.020 & -0.164 & $\pm$0.025 & -0.207 & $\pm$0.015 \\  
-3.5  & -0.218 &$\pm$0.010  & -0.008 & $\pm$0.020 & -0.166 & $\pm$0.025 & -0.222 & $\pm$0.015 \\
-2.9  & -0.228 &$\pm$0.010  &  0.002 & $\pm$0.021 & -0.167 & $\pm$0.025 & -0.236 & $\pm$0.015 \\
-2.2  & -0.240 &$\pm$0.010  &  0.013 & $\pm$0.021 & -0.165 & $\pm$0.026 & -0.251 & $\pm$0.015 \\
-1.5  & -0.253 &$\pm$0.010  &  0.025 & $\pm$0.021 & -0.162 & $\pm$0.026 & -0.265 & $\pm$0.016 \\
-0.8  & -0.266 &$\pm$0.010  &  0.036 & $\pm$0.020 & -0.155 & $\pm$0.026 & -0.279 & $\pm$0.016 \\
-0.1  & -0.280 &$\pm$0.010  &  0.046 & $\pm$0.019 & -0.146 & $\pm$0.026 & -0.291 & $\pm$0.016 \\
0.6   & -0.295 &$\pm$0.010  &  0.055 & $\pm$0.018 & -0.134 & $\pm$0.026 & -0.303 & $\pm$0.016 \\
1.3   & -0.310 &$\pm$0.010  &  0.061 & $\pm$0.018 & -0.119 & $\pm$0.026 & -0.314 & $\pm$0.016 \\
2.0   & -0.325 &$\pm$0.010  &  0.065 & $\pm$0.019 & -0.101 & $\pm$0.025 & -0.323 & $\pm$0.017 \\
2.7   & -0.340 &$\pm$0.010  &  0.064 & $\pm$0.020 & -0.080 & $\pm$0.025 & -0.330 & $\pm$0.017 \\
3.4   & -0.355 &$\pm$0.010  &  0.059 & $\pm$0.020 & -0.058 & $\pm$0.024 & -0.336 & $\pm$0.017 \\
4.1   & -0.369 &$\pm$0.011  &  0.049 & $\pm$0.020 & -0.033 & $\pm$0.024 & -0.340 & $\pm$0.017 \\
4.7   & -0.383 &$\pm$0.011  &  0.035 & $\pm$0.020 & -0.006 & $\pm$0.024 & -0.342 & $\pm$0.017 \\
5.4   & -0.396 &$\pm$0.011  &  0.017 & $\pm$0.020 &  0.022 & $\pm$0.024 & -0.343 & $\pm$0.018 \\
6.1   & -0.409 &$\pm$0.012  & -0.004 & $\pm$0.020 &  0.050 & $\pm$0.025 & -0.342 & $\pm$0.018 \\
6.8   & -0.422 &$\pm$0.012  & -0.028 & $\pm$0.020 &  0.079 & $\pm$0.026 & -0.339 & $\pm$0.018 \\
7.5   & -0.434 &$\pm$0.013  & -0.054 & $\pm$0.020 &  0.107 & $\pm$0.027 & -0.335 & $\pm$0.018 \\
8.2   & -0.446 &$\pm$0.014  & -0.082 & $\pm$0.021 &  0.130 & $\pm$0.028 & -0.329 & $\pm$0.019 \\
8.9   & -0.457 &$\pm$0.014  & -0.113 & $\pm$0.021 &  0.160 & $\pm$0.029 & -0.322 & $\pm$0.019 \\
9.6   & -0.468 &$\pm$0.015  & -0.145 & $\pm$0.021 &  0.190 & $\pm$0.030 & -0.314 & $\pm$0.020 \\
10.3  & -0.479 &$\pm$0.015  & -0.178 & $\pm$0.021 &  0.210 & $\pm$0.030 & -0.305 & $\pm$0.020 \\
11.0  & -0.488 &$\pm$0.015  & -0.212 & $\pm$0.022 &  0.230 & $\pm$0.030 & -0.295 & $\pm$0.020 \\
11.7  & -0.497 &$\pm$0.015  & -0.245 & $\pm$0.022 &  0.250 & $\pm$0.030 & -0.284 & $\pm$0.020 \\
12.3  & -0.505 &$\pm$0.014  & -0.277 & $\pm$0.023 &  0.270 & $\pm$0.029 & -0.273 & $\pm$0.019 \\
13.0  & -0.512 &$\pm$0.014  & -0.307 & $\pm$0.024 &  0.280 & $\pm$0.029 & -0.261 & $\pm$0.019 \\
13.7  & -0.517 &$\pm$0.014  & -0.334 & $\pm$0.024 &  0.300 & $\pm$0.028 & -0.249 & $\pm$0.019 \\
14.4  & -0.521 &$\pm$0.014  & -0.358 & $\pm$0.022 &  0.311 & $\pm$0.028 & -0.235 & $\pm$0.019 \\
15.1  & -0.523 &$\pm$0.014  & -0.379 & $\pm$0.021 &  0.322 & $\pm$0.028 & -0.221 & $\pm$0.018 \\
15.8  & -0.523 &$\pm$0.014  & -0.397 & $\pm$0.020 &  0.331 & $\pm$0.028 & -0.207 & $\pm$0.018 \\
16.5  & -0.521 &$\pm$0.014  & -0.413 & $\pm$0.020 &  0.340 & $\pm$0.028 & -0.193 & $\pm$0.018 \\
17.2  & -0.516 &$\pm$0.014  & -0.428 & $\pm$0.022 &  0.350 & $\pm$0.029 & -0.179 & $\pm$0.019 \\
17.9  & -0.509 &$\pm$0.014  & -0.442 & $\pm$0.023 &  0.350 & $\pm$0.030 & -0.165 & $\pm$0.020 \\
18.6  & -0.499 &$\pm$0.015  & -0.456 & $\pm$0.024 &  0.360 & $\pm$0.031 & -0.151 & $\pm$0.021 \\
19.2  & -0.487 &$\pm$0.015  & -0.468 & $\pm$0.025 &  0.360 & $\pm$0.033 & -0.139 & $\pm$0.021 \\
19.9  & -0.473 &$\pm$0.015  & -0.481 & $\pm$0.026 &  0.370 & $\pm$0.034 & -0.127 & $\pm$0.022 \\
20.6  & -0.456 &$\pm$0.015  & -0.493 & $\pm$0.027 &  0.370 & $\pm$0.036 & -0.116 & $\pm$0.022 \\
21.3  & -0.437 &$\pm$0.016  & -0.505 & $\pm$0.029 &  0.370 & $\pm$0.038 & -0.107 & $\pm$0.023 \\
22.0  & -0.416 &$\pm$0.016  & -0.518 & $\pm$0.031 &  0.380 & $\pm$0.039 & -0.098 & $\pm$0.024 \\
22.7  & -0.394 &$\pm$0.017  & -0.530 & $\pm$0.033 &  0.380 & $\pm$0.041 & -0.089 & $\pm$0.025 \\
23.4  & -0.370 &$\pm$0.017  & -0.543 & $\pm$0.034 &  0.380 & $\pm$0.042 & -0.082 & $\pm$0.026 \\
24.1  & -0.344 &$\pm$0.017  & -0.557 & $\pm$0.035 &  0.380 & $\pm$0.043 & -0.075 & $\pm$0.027 \\
24.8  & -0.318 &$\pm$0.018  & -0.572 & $\pm$0.036 &  0.380 & $\pm$0.043 & -0.068 & $\pm$0.029 \\
25.5  & -0.291 &$\pm$0.018  & -0.587 & $\pm$0.036 &  0.370 & $\pm$0.043 & -0.061 & $\pm$0.029 \\
26.2  & -0.264 &$\pm$0.018  & -0.604 & $\pm$0.035 &  0.370 & $\pm$0.043 & -0.055 & $\pm$0.030 \\
26.8  & -0.236 &$\pm$0.018  & -0.621 & $\pm$0.034 &  0.370 & $\pm$0.042 & -0.048 & $\pm$0.030 \\
27.5  & -0.209 &$\pm$0.018  & -0.639 & $\pm$0.034 &  0.370 & $\pm$0.042 & -0.041 & $\pm$0.030 \\
28.2  & -0.182 &$\pm$0.018  & -0.656 & $\pm$0.033 &  0.360 & $\pm$0.041 & -0.035 & $\pm$0.029 \\
28.9  & -0.155 &$\pm$0.018  & -0.672 & $\pm$0.033 &  0.360 & $\pm$0.041 & -0.028 & $\pm$0.028 \\
29.6  & -0.129 &$\pm$0.019  & -0.686 & $\pm$0.033 &  0.350 & $\pm$0.041 & -0.022 & $\pm$0.028 \\
30.3  & -0.105 &$\pm$0.019  & -0.699 & $\pm$0.034 &  0.350 & $\pm$0.041 & -0.016 & $\pm$0.028 \\
31.0  & -0.081 &$\pm$0.019  & -0.709 & $\pm$0.034 &  0.340 & $\pm$0.041 & -0.010 & $\pm$0.028 \\
31.7  & -0.059 &$\pm$0.020  & -0.717 & $\pm$0.036 &  0.330 & $\pm$0.042 & -0.005 & $\pm$0.029 \\
32.4  & -0.039 &$\pm$0.021  & -0.723 & $\pm$0.038 &  0.320 & $\pm$0.043 & -0.001 & $\pm$0.030 \\
33.1  & -0.019 &$\pm$0.022  & -0.726 & $\pm$0.041 &  0.310 & $\pm$0.045 &  0.002 & $\pm$0.031 \\
33.8  & -0.002 &$\pm$0.022  & -0.728 & $\pm$0.044 &  0.310 & $\pm$0.046 &  0.005 & $\pm$0.032 \\
34.4  &  0.014 &$\pm$0.023  & -0.729 & $\pm$0.048 &  0.300 & $\pm$0.048 &  0.007 & $\pm$0.033 \\
35.1  &  0.029 &$\pm$0.023  & -0.730 & $\pm$0.051 &  0.290 & $\pm$0.049 &  0.008 & $\pm$0.033 \\
35.8  &  0.042 &$\pm$0.023  & -0.732 & $\pm$0.052 &  0.280 & $\pm$0.049 &  0.009 & $\pm$0.033 \\
36.5  &  0.054 &$\pm$0.023  & -0.734 & $\pm$0.052 &  0.280 & $\pm$0.049 &  0.009 & $\pm$0.033 \\
37.2  &  0.064 &$\pm$0.023  & -0.738 & $\pm$0.050 &  0.270 & $\pm$0.048 &  0.008 & $\pm$0.032 \\
37.9  &  0.074 &$\pm$0.023  & -0.744 & $\pm$0.046 &  0.270 & $\pm$0.047 &  0.007 & $\pm$0.031 \\
38.6  &  0.082 &$\pm$0.023  & -0.751 & $\pm$0.042 &  0.270 & $\pm$0.045 &  0.006 & $\pm$0.030 \\
39.3  &  0.090 &$\pm$0.022  & -0.758 & $\pm$0.037 &  0.260 & $\pm$0.044 &  0.005 & $\pm$0.029 \\
40.0  &  0.096 &$\pm$0.022  & -0.765 & $\pm$0.035 &  0.260 & $\pm$0.042 &  0.004 & $\pm$0.029 \\
40.7  &  0.102 &$\pm$0.022  & -0.771 & $\pm$0.037 &  0.260 & $\pm$0.041 &  0.002 & $\pm$0.028 \\
41.4  &  0.107 &$\pm$0.022  & -0.773 & $\pm$0.040 &  0.260 & $\pm$0.040 &  0.001 & $\pm$0.028 \\
42.0  &  0.112 &$\pm$0.022  & -0.772 & $\pm$0.044 &  0.260 & $\pm$0.039 & -0.001 & $\pm$0.028 \\
42.7  &  0.116 &$\pm$0.022  & -0.767 & $\pm$0.045 &  0.260 & $\pm$0.039 & -0.003 & $\pm$0.028 \\
43.4  &  0.120 &$\pm$0.022  & -0.756 & $\pm$0.044 &  0.260 & $\pm$0.039 & -0.005 & $\pm$0.028 \\
44.1  &  0.124 &$\pm$0.023  & -0.742 & $\pm$0.041 &  0.260 & $\pm$0.040 & -0.006 & $\pm$0.029 \\
44.8  &  0.128 &$\pm$0.023  & -0.724 & $\pm$0.038 &  0.260 & $\pm$0.041 & -0.007 & $\pm$0.029 \\
45.5  &  0.132 &$\pm$0.024  & -0.704 & $\pm$0.037 &  0.260 & $\pm$0.042 & -0.008 & $\pm$0.030 \\
46.2  &  0.137 &$\pm$0.025  & -0.684 & $\pm$0.041 &  0.260 & $\pm$0.044 & -0.008 & $\pm$0.031 \\
46.9  &  0.141 &$\pm$0.025  & -0.665 & $\pm$0.049 &  0.260 & $\pm$0.046 & -0.007 & $\pm$0.032 \\
47.6  &  0.147 &$\pm$0.026  & -0.650 & $\pm$0.059 &  0.260 & $\pm$0.048 & -0.006 & $\pm$0.033 \\
48.3  &  0.153 &$\pm$0.027  & -0.639 & $\pm$0.067 &  0.260 & $\pm$0.050 & -0.005 & $\pm$0.034 \\
49.0  &  0.159 &$\pm$0.028  & -0.635 & $\pm$0.072 &  0.260 & $\pm$0.053 & -0.003 & $\pm$0.035 \\
49.6  &  0.166 &$\pm$0.029  & -0.638 & $\pm$0.073 &  0.260 & $\pm$0.056 & -0.002 & $\pm$0.037 \\
50.3  &  0.173 &$\pm$0.029  & -0.648 & $\pm$0.071 &  0.260 & $\pm$0.059 & -0.000 & $\pm$0.039 \\
51.0  &  0.180 &$\pm$0.030  & -0.666 & $\pm$0.066 &  0.260 & $\pm$0.062 &  0.001 & $\pm$0.041 \\
51.7  &  0.187 &$\pm$0.032  & -0.689 & $\pm$0.059 &  0.270 & $\pm$0.065 &  0.002 & $\pm$0.044 \\
52.4  &  0.194 &$\pm$0.033  & -0.718 & $\pm$0.054 &  0.270 & $\pm$0.069 &  0.003 & $\pm$0.046 \\
53.1  &  0.201 &$\pm$0.036  & -0.751 & $\pm$0.054 &  0.270 & $\pm$0.074 &  0.004 & $\pm$0.050 \\
53.8  &  0.208 &$\pm$0.039  & -0.787 & $\pm$0.062 &  0.270 & $\pm$0.079 &  0.004 & $\pm$0.054 \\
\enddata
\tablecomments{We reiterate that these $K$~correction values are provided only for reference,
 and {\it should not} be blindly used to correct NIR photometry for any given SN~Ia. }
\end{deluxetable}

\clearpage
\begin{deluxetable}{ccccc}
\tablewidth{0pt}
\tablecolumns{5}
\tablenum{3}
\tablewidth{0pt}
\tablecaption{MCMC values of $\sigma_{var}$\label{table3}}
\tablehead{
\colhead{Red-shift}  &
\colhead{$Y$}   &
\colhead{$J$}   &
\colhead{$H$}   &
\colhead{$K_s$}}
\startdata
0.030 & 0.015 & 0.027 & 0.020    & 0.029 \\  
0.050 & 0.024 & 0.038 & 0.057    & 0.045 \\   
0.080 & 0.038 & 0.058 & 0.095    & 0.065 \\   
\enddata
\end{deluxetable}

\end{document}